\documentclass{article}
\usepackage{geometry}
\usepackage{amsmath}
\usepackage{amsfonts}
\usepackage{amstext}
\usepackage{amssymb}
\usepackage{amsthm}
\usepackage{eufrak}
\newtheorem{lemma}{Lemma}

\newtheorem{theorem}{Theorem}
\newtheorem{corollary}{Corollary}
\theoremstyle{definition}
\newtheorem{definition}{Definition}

\theoremstyle{remark}
\newtheorem{remark}{Remark}

\usepackage{color}
\usepackage{graphicx} 
\begin{document}
\begin{center}
 \large \bf{Asymptotic analysis of stock price densities and implied volatilities in mixed stochastic models}
\end{center}
\vspace{0.5cm}
\begin{center}
\bf Archil Gulisashvili\,$^a$,\quad Josep Vives\,$^b$ \rm
\end{center}
\vspace{0.1in}
\begin{center}
\scriptsize\it $^a$\,Department of Mathematics, Ohio University, Athens, OH 45701, USA
\\
$^b$\,Departament de Probabilitat, L\`{o}gica i Estad\'{i}stica, Universitat de Barcelona,
08007-Barcelona (Catalunya), Spain \normalsize\rm
\end{center}
 


\begin{abstract}
In this paper, we obtain sharp asymptotic formulas with error estimates for the Mellin convolution of functions defined on $(0,\infty)$,
and use these formulas to characterize the asymptotic behavior of marginal distribution densities of stock price processes in mixed stochastic models. Special examples of mixed models are jump-diffusion models and stochastic volatility models with jumps. We apply our general results to the Heston model with double exponential jumps, and make a detailed analysis of the asymptotic behavior of the stock price density, the call option pricing function, and the implied volatility in this model. We also obtain similar results for the Heston model with jumps distributed according to the NIG law. 
\end{abstract}

\section{Introduction}\label{S:int}
The random behavior of the stock price in a mixed model is described by a stochastic process $X=X^{(1)}X^{(2)}$, where $X^{(1)}$ and $X^{(2)}$ are strictly positive independent integrable processes on a complete filtered probability space $(\Omega,{\cal F},\{{\cal F}_t\},\mathbb{P})$. Important examples of mixed models are jump-diffusion models and stochastic volatility models with L\'{e}vy type jumps. More information on models with jumps
can be found in \cite{CT} and \cite{S}. 

In this paper, we obtain asymptotic formulas with error estimates for the distribution density of the stock price and the implied volatility in special mixed stochastic stock price models. Let us suppose that the distributions $\mu^{(1)}_t$ and $\mu^{(2)}_t$ of the random variables $X^{(1)}_t$ and $X^{(2)}_t$ in a mixed stochastic model admit densities $D^{(1)}_t$ and $D^{(2)}_t$, respectively. Then 
the distribution $\mu_t$ of the stock price $X_t$ also admits density $D_t$, which can be represented by the Mellin convolution 
\begin{equation}
D_t(x)=D^{(1)}_t\stackrel{M}{\star} D^{(2)}_t(x),\quad x> 0
\label{E:nach}
\end{equation} 
(the definition of the Mellin convolution is given below). The fact that the distribution density of the product of two independent random variables is the Mellin convolution of their densities was mentioned in \cite{E}.

In \cite{A} (see also \cite{BGT}), D. Arandelovi\'{c} obtained an asymptotic formula for the Mellin convolution of functions defined on the half-line 
$(0,\infty)$. However, Arandelovi\'{c}'s formula does not contain an 
\\
------------------------------------------------------------------------------------------------------------------------
 \\ 
\it E-mail addresses: \rm gulisash@ohio.edu (A. Gulisashvili),\,\,josep.vives@ub.edu 
(J. Vives).
\\
\normalsize
error estimate. In Subsection \ref{SS:Aranda} of the present paper, asymptotic formulas with error estimates are established for the Mellin convolution (see the formulas in Theorems \ref{T:Atg} and \ref{T:MC0}). These formulas extend Arandelovi\'{c}'s result. They are used in the paper to characterize the asymptotic behavior of the stock price density in special mixed stochastic stock price models. Note that asymptotic expansions of the Mellin convolution under different restictions than those imposed in Arandelovi\'{c}'s work and in the present paper were obtained by R. A. Handelsman and J. S. Lew 
(see the presentation of their results in Section 3.4 of \cite{W}).
 
One of the examples considered below is the Heston model with asymmetric double exponential jumps. Theorems 
\ref{T:glav} and \ref{T:nearze} obtained in this paper deal with the case where the jump part of the mixed model dominates, while Theorems \ref{T:glavk} and \ref{T:nearzek} 
concern the asymptotics of the stock price density in the case where the Heston part dominates. Weaker estimates were obtained earlier in \cite{GV}. In Section \ref{S:moor}, we briefly discuss some other models.

In \cite{K} (see also \cite{KW}), S. Kou introduced and studied a jump-diffusion model that is in fact a mixture of the Black-Scholes model with the double exponential jump model. An asymptotic formula (without an error estimate) for the distribution function of the stock price in the Kou model 
was obtained in \cite{ASu}, Example 7.6. In \cite{Z} and \cite{GMZ}, an asymptotic formula with an error estimate was found for the call pricing function in the Kou model. In the present paper, we obtain a similar formula for a class of models, including the Heston model with double exponential jumps and the Kou model (see  (\ref{E:si})). The error estimate in formula (\ref{E:si}) is better than that in \cite{GMZ}.

In our analysis of the stock price density in the Heston model perturbed by double exponential jumps, we use some of the results obtained in \cite{K}. It is interesting to mention that the asymptotic behavior of the stock price density in the Heston model without jumps and that of the absolutely continuous part of the distribution associated with the double exponential jump part is similar (compare (\ref{E:0S}) with (\ref{E:ur1}) and (\ref{E:estS}) with (\ref{E:ur2})). It follows that in the study of the asymptotic behavior of the stock price density in the mixture of the Heston model with the double exponential jump model, we have to take into account which part of the mixed model dominates the other. This dichotomy does not appear in the Kou model since the double exponential jump part always dominates the Black-Scholes part. Note that the similarity between the asymptotic behavior of the call pricing functions in the Heston model without jumps and in the Kou model was 
observed in \cite{GMZ} too.

Asymptotic formulas for the stock price density can be used to study the asymptotic behavior of option pricing functions and the implied volatility. In Section \ref{S:Hemj} of the present paper, we obtain asymptotic formulas with five explicit terms and error estimates for the implied volatility at extreme strikes in the Heston model with double exponential jumps. Similar formulas for the Heston model with NIG type jumps are established in Section \ref{S:moor}. We use some of the methods developed in \cite{GL} and \cite{G2} to estimate the implied volatility. A little weaker asymptotic formulas for the implied volatility with four explicit terms were established for the Heston model without jumps in \cite{GL} and for the Kou model in \cite{Z} and \cite{GMZ}. 
These formulas can be extended to include five terms and an error estimate. We would also like to bring the reader's attention to the paper \cite{AL} concerning the asymptotic behavior of the implied volatility in exponential L\'{e}vy models.

We will next briefly overview the contents of the present paper. In Subsection \ref{SS:MM}, we define the Mellin convolution and introduce several related notions. Regularly varying functions play an important role in the paper. In Subsection \ref{SS:rvf}, various definitions and facts from the theory of regularly varying functions are gathered, while Subsection \ref{SS:Aranda} is devoted to Arandelovi\'{c}'s theorem and its generalizations. In Section \ref{S:Hmod}, known asymptotic formulas for marginal distribution densities of the stock price process in the Heston model are formulated. Section \ref{S:ku} is devoted to the Heston model with double exponential jumps. Here we obtain new results concerning the jump part of the perturbed Heston model. We prove that the absolutely continuous part of the marginal distribution of the exponential L\'{e}vy process associated with the perturbed Heston model is regularly varying, and provide asymptotic formulas characterizing its asymptotic behavior near infinity and near zero.
Section \ref{S:deg} of the present paper deals with density approximations in the Heston model with double exponential jumps. We obtain sharp asymptotic formulas with error estimates for the distribution density of the stock price in the perturbed Heston model (see Theorems \ref{T:glav}-\ref{T:nearzek}). The generalizations of Arandelovi\'{c}'s theorem obtained in Theorems \ref{T:Atg} and \ref{T:MC0} are used in the proofs. In Section \ref{S:Hemj}, sharp asymptotic formulas with error estimates are provided for the implied volatility in the Heston model with double exponential jumps (see Theorems \ref{T:hdej} and \ref{T:kon}) and for more general models. 
Finally, Section \ref{S:moor} discusses the implied volatility in the Heston model with jumps distributed according to the symmetric centered NIG law.

\section{The Mellin convolution and Arandelovi\'{c}'s theorem}\label{S:MCAT}
In this section, we discuss the Mellin transform and the Mellin convolution, formulate Arandelovi\'{c} theorem,
concerning the asymptotics of the Mellin convolution, and obtain generalizations of Arandelovi\'{c}'s theorem
(Theorems \ref{T:Atg} and \ref{T:MC0}). 
\subsection{The Mellin transform and the Mellin convolution}\label{SS:MM}
\begin{definition}\label{D:Mellin1}
Let $U$ be a measurable function on $(0,\infty).$ The Mellin transform of $U$ is defined 
as follows: 

\begin{equation}\label{Mellin1}
{MU}(z)=\int_0^{\infty} t^{-z} U(t) \frac{dt}{t},\quad z\in\mathbb{C}.
\end{equation}
\end{definition}

The domain of the Mellin transform of $U$ is the set of all $z\in {\mathbb C}$ for which the integral in 
(\ref{Mellin1}) converges absolutely. 
\begin{definition}\label{D:MC}
The Mellin convolution of two real Lebesgue measurable functions $f$ and $g$ on $(0,\infty)$ is defined by 
$$  
f\stackrel{M}{\star}g(x)=\int_0^{\infty} f(t^{-1}x)g(t)\frac{dt}{t},
$$
for those $x>0$ for which the integral exists.
\end{definition}
It is clear that
$$
f\stackrel{M}{\star}g(x)=\int_0^{\infty} f(t^{-1})g(xt)\frac{dt}{t}.
$$
Moreover,
\begin{equation}
f\stackrel{M}{\star}g(x)=\widetilde{f}\stackrel{M}{\star}\widetilde{g}(x^{-1}),
\label{E:mel}
\end{equation}
where 
\begin{equation}
\widetilde{f}(u)=f(u^{-1})\quad\mbox{and}\quad \widetilde{g}(u)=g(u^{-1})
\label{E:inv}
\end{equation}
for all $u> 0$. We also have
\begin{equation}
M\widetilde{U}(z)=MU(-z).
\label{E:min}
\end{equation}

Let $\mu$ be a distribution on $[0,\infty)$, and let $\eta$ be a real number. The moment of order $\eta$ of the distribution $\mu$ is defined as follows:
\begin{equation}
m_{\eta}(\mu)=\int_0^{\infty}t^{\eta}d\mu(t).
\label{E:moment}
\end{equation}
It is not hard to see that if $U$ is a distribution density, then
$MU(\eta)=m_{-\eta-1}(U)$ for all real numbers $\eta$ in the domain of $MU$.

\subsection{Regularly varying functions}\label{SS:rvf}
In the present subsection, several notions and results from the theory of regularly varying functions are discussed. These functions play an important role in the paper. A rich source of information about regularly varying functions is the book \cite{BGT} by N. H. Bingham, C. M. Goldie, and J. L. Teugels. 
\begin{definition}
A nonnegative measurable function $f$ on $(0,\infty)$ is called regularly varying with index $\rho\in {\mathbb R}$ if 
for every $\lambda>0,$ 
\begin{equation}
\frac{f(\lambda x)}{f(x)}\rightarrow \lambda^{\rho}
\label{E:c1}
\end{equation}
as $x\rightarrow \infty.$
The class of all regularly varying functions with index $\rho$ is denoted by $R_{\rho}.$ Functions from the class $R_0$ are called slowly varying functions. 
\end{definition}

The next result is known as the uniform convergence theorem for regularly varying functions.
\begin{theorem}[see Theorem 1.5.2 in \cite{BGT}]\label{T:UCT1}
Let $f$ be a nonnegative measurable function on $(0,\infty)$. Then the following are true:
\begin{enumerate} 
\item Suppose $f\in R_{\rho}$ with $\rho> 0$ and $f$ is bounded on every interval
$(0,a]$ with $a> 0$. Then formula (\ref{E:c1}) holds uniformly in $\lambda$ on each interval $(0,a]$, $a> 0$.
\item The condition $f\in R_0$ implies that formula (\ref{E:c1}) holds uniformly in $\lambda$ on each interval
$[a,b]$ with $0<a< b<\infty$.
\item The condition $f\in R_{\rho}$ with $\rho< 0$ implies that formula (\ref{E:c1}) holds uniformly in $\lambda$ on each interval $[b,\infty)$, $b> 0$.
\end{enumerate}
\end{theorem}

Another fundamental result in the theory of slowly varying functions is the representation theorem (see \cite{BGT}, Theorem 1.3.1).
\begin{theorem}\label{T:RT}
For a nonnegative measurable function $l$, the condition $l\in R_0$ is equivalent to the following:
\begin{equation}
l(x)=c(x)\exp\left\{\int_a^x\frac{\varepsilon(u)}{u}du\right\},\quad x> a,
\label{E:lx}
\end{equation}
for some $a> 0$, where the functions $c$ and $\varepsilon$ are such that $c(x)\rightarrow c\in(0,\infty)$ as $x\rightarrow\infty$ 
and $\varepsilon(u)\rightarrow 0$ as $u\rightarrow\infty$.
\end{theorem}
\begin{definition}\label{D:NSV}
A function $l\in R_0$ is called a normalized slowly varying function provided that the function $x\mapsto c(x)$ in 
(\ref{E:lx}) is constant on the interval $[a,\infty)$ for some $a> 0$.
\end{definition}

Let $l$ be a normalized slowly varying function. Then Theorem \ref{T:RT} shows that 
\begin{equation}
l(x)=\exp\left\{C+\int_a^x\frac{\varepsilon(u)}{u}du\right\},\quad x> a,
\label{E:tsh}
\end{equation}
for some $C\in\mathbb{R}$ and $a> 0$, where the function $\varepsilon$ is such that $\varepsilon(u)\rightarrow 0$ as $u\rightarrow\infty$.

For a normalized slowly varying function $l$, we have
\begin{equation}
\varepsilon(x)=\frac{xl^{\prime}(x)}{l(x)}\quad\mbox{a.e.}
\label{E:forn}
\end{equation}
(see \cite{BGT}, p.\,15). If the function $l$ is differentiable, then the equality in (\ref{E:forn})
holds everywhere on $(a,\infty)$.

The following class was introduced by A. Zygmund.
\begin{definition}\label{D:Zy}
A nonnegative measurable function $l$ defined on $(0,\infty)$ belongs to the Zygmund class $\cal Z$ if, for every $\alpha>0,$ the function 
$\phi_{\alpha}(x)=x^{\alpha} l(x)$ is ultimately 
increasing and the function 
$\psi_{\alpha}(x)=x^{-\alpha} l(x)$
is ultimately decreasing. 
\end{definition}

The next known statement gives a description of the Zygmund class.
\begin{theorem}[see Theorem 1.5.5. in \cite{BGT}]\label{T:NSWZ}
The class $\cal Z$ coincides with the class of normalized slowly varying functions. 
\end{theorem}

In the present paper, we discuss various asymptotic formulas with error estimates. Note that throughout the paper 
the statement $\phi_1(x)=O(\phi_2(x))$ as $x\rightarrow\infty$, where $\phi_1$ is a real function and $\phi_2$ is a positive function, means that there exist $c> 0$ and $x_0> 0$ such that $|\phi_1(x)|\le c\phi_2(x)$ for all $x> x_0$. The explanation of the statement $\phi_1(x)=O(\phi_2(x))$ as $x\rightarrow 0$ is similar.

The following known definition introduces slowly varying functions with remainder (see \cite{GoS}, 
see also \cite{BGT}).
\begin{definition}\label{D:RR}
Let $l$ and $g$ be nonnegative measurable functions on $(0,\infty)$ with $g(x)\rightarrow 0$ as $x\rightarrow\infty$. The function $l$ is called slowly varying with remainder $g$
if for all $\lambda> 1$,
\begin{equation}
\frac{l(\lambda x)}{l(x)}-1=O(g(x))
\label{E:u1}
\end{equation}
as $x\rightarrow\infty$.
\end{definition}

We will denote the class of slowly varying functions with remainder $g$ by $R_0^g$. It is not hard to see that
$R_0^g\subset R_0$. The uniform convergence theorem for slowly varying functions with 
remainder is as follows.
\begin{theorem}[see Corollary 2.2.1 in \cite{BGT}]\label{T:UCT2}
Let $l\in R_0^g$ where $g\in R_0$. Then condition (\ref{E:u1}) holds uniformly in $\lambda$ on every interval
$[1,b]$, $b> 1$.
\end{theorem}

The next statement, which is stronger than Theorem \ref{T:UCT2}, provides a growth estimate in the variable $\lambda$ in the uniform convergence result for slowly varying functions with remainder.
\begin{theorem}\label{T:zabb}
Fix $\delta\neq 0$, and let $f$ and $g$ be positive functions on $[0,\infty)$ such that $g\in R_0$ and $f\in R_0^g$.
Suppose also that the functions $f$, $g$, $\frac{1}{f}$, and $\frac{1}{g}$ are locally bounded on
$[0,\infty)$. Then there exists $A> 0$ such that
$$
|f(\lambda x)-f(x)|\le Af(x)g(x)\max\{\lambda^{\delta},\lambda^{-\delta}\}
$$
for all $\lambda> 0$ and $x\ge 0$.
\end{theorem}

The estimate in Theorem \ref{T:zabb} is contained in part (b) of Theorem 3.8.6 in \cite{BGT}. Note that the condition $f\in R_0^g$
implies the following inclusion: $f\in O\Pi_l$ where $l=fg$, and hence the conditions in Theorem 3.8.6 (b)
hold (see \cite{BGT} for the definition of the class $O\Pi_l$ and for more details).

The structure of slowly varying functions with remainder is known. The next result is the representation theorem
for slowly varying functions with remainder (see \cite{GoS}, see also \cite{BGT}).
\begin{theorem}\label{T:GS}
Let $g\in R_0$ and $g(x)\rightarrow 0$ as $x\rightarrow\infty$. Then $l\in R_0^g$ if and only if
\begin{equation}
l(x)=\exp\left\{C+O(g(x))+\int_a^xO(g(t))t^{-1}dt\right\}
\label{E:ops}
\end{equation}
as $x\rightarrow\infty$, where $C\in\mathbb{R}$, and the $O$ functions are locally integrable.
\end{theorem}
\begin{corollary}\label{C:use}
Let $l\in Z$. Then $l\in R_0^{|\varepsilon|}$, where $\varepsilon$ is the function appearing in formula
(\ref{E:tsh}).
\end{corollary}

\it Proof. \rm Since $l$ is a normalized slowly varying function, formula (\ref{E:tsh}) holds. This implies that formula (\ref{E:ops}) holds with $g=|\varepsilon|$. Next, using Theorem \ref{T:GS}, we establish Corollary 
\ref{C:use}.
\subsection{Arandelovic's theorem and its generalizations}\label{SS:Aranda}
The next statement was obtained by Arandelovi\'{c} (see \cite{A,BGT}).
\begin{theorem}\label{T:arand}
Suppose the Mellin transform $MU$ of the function $U$ converges at least in the strip $\sigma\leq \Re(z) \leq \tau$ where $-\infty<\sigma<\tau<\infty.$
Let $f$ be a measurable function on $(0,\infty),$ and assume the following two conditions hold: 

\begin{enumerate}

\item
$f(x)\sim x^{\rho} l(x)$ as $x\rightarrow \infty$ where $\rho\in (\sigma,\tau)$ and $l\in R_0.$

\item
The function $x\mapsto x^{-\sigma}f(x)$ is bounded on every interval $(0,a]$ where $a>0.$
\end{enumerate}

Then 

\begin{equation}\label{Mellin3}
U\stackrel{M}{\star}f(x)\sim {MU}(\rho)[x^{\rho}l(x)]
\end{equation}
as $x\rightarrow \infty.$
\end{theorem}

Note that there is no error estimate in formula (\ref{Mellin3}). The next assertion provides such an estimate under certain additional restrictions.
\begin{theorem}\label{T:Atg}
Suppose the Mellin transform $MU$ of a measurable function $U$ converges at least in the strip $\sigma\leq \Re(z) \leq \tau$ where $-\infty<\sigma<\tau<\infty.$ Let $f$ be a measurable function on $(0,\infty),$ and assume the following  conditions hold: 
\begin{enumerate}

\item
$f(x)=x^{\rho}l(x)(1+O(h(x)))$ as $x\rightarrow \infty$ where $\rho\in (\sigma,\tau)$, $l\in R_0^g$ with $g\in R_0$,
and $h\in Z$. The functions $g$ and $h$ in the previous formula satisfy $g(x)\rightarrow 0$ and $h(x)\rightarrow 0$ as
$x\rightarrow\infty$.

\item The functions $g$, $l^{-1}$, and $g^{-1}$ are locally bounded on the interval $(x_0,\infty)$ for some $x_0> 0$. 

\item
The function $x\mapsto x^{-\sigma}f(x)$ is bounded on every interval $(0,a]$ where $a>0.$
\end{enumerate}
Then 

\begin{equation}\label{Mellins}
U\stackrel{M}{\star}f(x)={MU}(\rho)[x^{\rho}l(x)](1+O(g(x))+O(h(x)))
\end{equation}
as $x\rightarrow \infty.$

\end{theorem}

\it Proof. \rm Let us first assume that 
\begin{equation}
f(y)=y^{\rho}l(y)\quad\mbox{for}\quad y> y_0,
\label{E:adc}
\end{equation}
and put $x_1=\max\{x_0,y_0\}$, where $x_0$ is such as in condition 2. Then we have
\begin{align}
U\stackrel{M}{\star}f(x)&=\int_0^{\frac{x_1}{x}}f(xv)U(v^{-1})\frac{dv}{v}
+x^{\rho}\int_{\frac{x_1}{x}}^{\infty}v^{\rho}l(xv)U(v^{-1})\frac{dv}{v} \nonumber \\
&=I_1(x)+I_2(x).
\label{E:kor2}
\end{align}

Since $xv< x_1$ in the first integral in (\ref{E:kor2}), condition 3 implies the following:
\begin{align*}
I_1(x)&\le cx^{\sigma}\int_0^{\frac{x_1}{x}}v^{\sigma}U(v^{-1})\frac{dv}{v}= cx^{\sigma}\int_{\frac{x}{x_1}}
^{\infty}y^{-\sigma-1}U(y)dy \\
&\le cMU(\sigma)x^{\sigma}.
\end{align*}
Fix $\varepsilon> 0$ such that $\varepsilon<\rho-\sigma$, and take into account that $lg\in R_0$. Then the previous estimates imply that
$I_1(x)=O(x^{\sigma})=O(x^{\rho-\varepsilon})$, and hence
\begin{equation}
I_1(x)=O(x^{\rho}l(x)g(x))\quad\mbox{as}\quad x\rightarrow\infty.
\label{E:kor3}
\end{equation}
It follows that the integral $I_1$ can be incorporated into the error term in formula (\ref{Mellins}).

Our next goal is to estimate the integral $I_2$. It is clear that
\begin{align}
I_2(x)&=x^{\rho}\int_{\frac{x_1}{x}}^{\infty}[l(xv)-l(x)]v^{\rho}U(v^{-1})\frac{dv}{v}
+x^{\rho}l(x)\int_{\frac{x_1}{x}}^{\infty}v^{\rho}U(v^{-1})\frac{dv}{v} \nonumber \\
&=x^{\rho}l(x)\int_0^{\infty}v^{\rho}U(v^{-1})\frac{dv}{v}-x^{\rho}l(x)
\int_0^{\frac{x_1}{x}}v^{\rho}U(v^{-1})\frac{dv}{v} \nonumber \\
&\quad+x^{\rho}\int_{\frac{x_1}{x}}^{\infty}[l(xv)-l(x)]v^{\rho}U(v^{-1})\frac{dv}{v} \nonumber \\
&=MU(\rho)[x^{\rho}l(x)]-x^{\rho}l(x)
\int_{\frac{x}{x_1}}^{\infty}y^{-\rho}U(y)\frac{dy}{y} \nonumber \\
&\quad+x^{\rho}\int_{\frac{x_1}{x}}^{\infty}[l(xv)-l(x)]v^{\rho}U(v^{-1})\frac{dv}{v} \nonumber \\
&=MU(\rho)[x^{\rho}l(x)]+J_1(x)+J_2(x).
\label{E:kor4}
\end{align}

Fix $\varepsilon> 0$ such that $\varepsilon<\rho-\sigma$. Then, for large values of $x$, we have
$$
|J_1(x)|\le x^{\rho}l(x)\int_{\frac{x}{x_1}}^{\infty}y^{-\sigma-\varepsilon}U(y)\frac{dy}{y}
\le x^{\rho-\varepsilon}l(x)x_1^{\varepsilon}MU(\sigma).
$$
Since $g\in R_0$ and $MU(\sigma)<\infty$, we obtain
\begin{equation}
J_1(x)=O(x^{\rho}l(x)g(x))\quad\mbox{as}\quad x\rightarrow\infty.
\label{E:kor5}
\end{equation}

It remains to estimate $J_2$. Denote by $\widetilde{l}$ and $\widetilde{g}$ the functions $l$ and $g$,
extrapolated by positive constants from $(x_1,\infty)$ to $[0,\infty)$. Then $\widetilde{l}\in R_0^{\widetilde{g}}$.
Moreover, condition 2 and the definition of the functions $\widetilde{l}$ and $\widetilde{g}$
imply that the functions $\widetilde{l}$, $\widetilde{g}$, $\widetilde{l}^{-1}$, and $\widetilde{g}^{-1}$ are locally bounded on $[0,\infty)$. Now, using Theorem \ref{T:zabb}, we see that for every $\delta> 0$ 
there exists a constant $A> 0$ such that
$$
|\widetilde{l}(xv)-\widetilde{l}(x)|\le A\widetilde{l}(x)\widetilde{g}(x)\max\{v^{\delta},v^{-\delta}\}
$$
for all $v> 0$ and $x\ge 0$. Recalling the definition of the functions $\widetilde{l}$ and $\widetilde{g}$,
we see that for $x> x_1$ and $v>\frac{x_1}{x}$,
\begin{equation}
|l(xv)-l(x)|\le Al(x)g(x)\max\{v^{\delta},v^{-\delta}\}.
\label{E:kor6}
\end{equation}
It follows from the estimate in (\ref{E:kor6}) that for every $\delta> 0$ there exists $x_1> 0$ depending on $\delta$
and such that
\begin{align}
|J_2(x)|&\le Ax^{\rho}l(x)g(x)\int_{\frac{x_1}{x}}^{\infty}\max\{v^{\delta},v^{-\delta}\}
v^{\rho}U(v^{-1})\frac{dv}{v} \nonumber \\
&\le Ax^{\rho}l(x)g(x)\int_0^{\infty}\max\{v^{\delta},v^{-\delta}\}
v^{\rho}U(v^{-1})\frac{dv}{v},
\label{E:kor7}
\end{align}
for all $x> x_1$. It is not hard to see that for small enough values of $\delta$, the last integral in 
(\ref{E:kor7}) is finite. Here we use the fact that $MU(s)<\infty$ for all $\sigma\le s\le\tau$ and the inequalities
$\sigma<\rho<\tau$. Now, (\ref{E:kor7}) implies that
\begin{equation}
J_2(x)=O(x^{\rho}l(x)g(x))\quad\mbox{as}\quad x\rightarrow\infty.
\label{E:kor8}
\end{equation}
Next, taking into account formulas (\ref{E:kor4}), (\ref{E:kor5}), and (\ref{E:kor8}), we obtain
\begin{equation}
I_2(x)=MU(\rho)[x^{\rho}l(x)]+O(x^{\rho}l(x)g(x))
\label{E:fine}
\end{equation}
as $x\rightarrow\infty$. Finally, it is easy to see that formulas (\ref{E:kor2}), (\ref{E:kor3}), and (\ref{E:fine})
imply
formula (\ref{Mellins}). This establishes Theorem \ref{T:Atg} in a special case where (\ref{E:adc})
holds.

We will next prove Theorem \ref{T:Atg} in the general case. Suppose the conditions in the formulation of Theorem
\ref{T:Atg} hold. Then there exists $x_1> 0$ such that 
$f(x)=x^{\rho}l(x)+x^{\rho}l(x)\eta(x)$
for all $x> x_1$, where $\eta$ is a measurable function such that 
$|\eta(x)|\le Ah(x)$,
for all $x> x_1$ and for some constant $A> 0$. Put $f_1(x)=f(x)\chi_{\{0< x< x_1\}}$, 
$f_2(x)=x^{\rho}l(x)\chi_{\{x> x_1\}}$, and $f_3(x)=x^{\rho}l(x)\eta(x)\chi_{\{x> x_1\}}$.
Then
\begin{equation}
U\stackrel{M}{\star}f(x)=U\stackrel{M}{\star}f_1(x)+U\stackrel{M}{\star}f_2(x)+U\stackrel{M}{\star}f_3(x).
\label{E:m1}
\end{equation}

Applying the special case of Theorem \ref{T:Atg} established above to the function $f_2$, we obtain
\begin{equation}
U\stackrel{M}{\star}f_2(x)={MU}(\rho)[x^{\rho}l(x)](1+O(g(x)))
\label{E:rrr1}
\end{equation}
as $x\rightarrow\infty$. In addition, reasoning as in the proof of (\ref{E:kor3}), we obtain
\begin{equation}
U\stackrel{M}{\star}f_1(x)=O(x^{\rho}l(x)g(x))
\label{E:rrr2}
\end{equation}
as $x\rightarrow\infty$.

We will next estimate the function $U\stackrel{M}{\star}f_3$, using the same ideas as in the estimate for the function $U\stackrel{M}{\star}f_2$. However, we will use the function $\widetilde{l}=lh$ instead of the function $l$. Recall that
$l\in R_0^g$. Moreover, $h\in R_0^{|\varepsilon|}$, where $\varepsilon$ is the function appearing in formula (\ref{E:tsh}) for the function $h$ (see Corollary \ref{C:use}).

We have 
$$
\left|\frac{\widetilde{l}(xv)}{\widetilde{l}(x)}-1\right|\le\left|\frac{l(xv)}{l(x)}-1\right|\frac{h(xv)}{h(x)}
+\left|\frac{h(xv)}{h(x)}-1\right|.
$$
Therefore, $\widetilde{l}\in R_0^{g+|\varepsilon|}$. It is clear that
$$
\left|U\stackrel{M}{\star}f_3(x)\right|\le Ax^{\rho}\int_{\frac{x_1}{x}}^{\infty}v^{\rho}l(xv)h(xv)U(v^{-1})
\frac{dv}{v}.
$$
Now, reasoning as in the proof of formula (\ref{E:fine}), we see that
\begin{align}
U\stackrel{M}{\star}f_3(x)&=O\left(x^{\rho}l(x)[h(x)+h(x)g(x)+h(x)|\varepsilon(x)|]\right) \nonumber \\
&=O\left(x^{\rho}l(x)h(x)\right)
\label{E:rrr3}
\end{align}
as $x\rightarrow\infty$.

Finally, taking into account formulas (\ref{E:m1}) - (\ref{E:rrr3}), we see that formula (\ref{Mellins})
holds. This completes the proof of Theorem \ref{T:Atg}.

A similar theorem characterizes the asymptotic behavior of the Mellin convolution near zero.
\begin{theorem}\label{T:MC0}
Suppose the Mellin transform $MU$ of a measurable function $U$ converges at least in the strip $\sigma\leq \Re(z) \leq \tau$ where $-\infty<\sigma<\tau<\infty.$ Let $f$ be a measurable function on $(0,\infty),$ and assume the following  conditions hold: 
\begin{enumerate}

\item
$f(y^{-1})=y^{-\rho}l(y)(1+O(h(y)))$ as $y\rightarrow \infty$ where $\rho\in(\sigma,\tau)$, $l\in R_0^g$ with $g\in R_0$,
and $h\in Z$. The functions $g$ and $h$ in the previous formula satisfy $g(y)\rightarrow 0$ and $h(y)\rightarrow 0$ as
$y\rightarrow\infty$.

\item The functions $g$, $l^{-1}$, and $g^{-1}$ are locally bounded on the interval $(x_0,\infty)$ for some $x_0> 0$. 

\item
The function $y\mapsto y^{\tau}f(y^{-1})$ is bounded on every interval $(0,a]$ where $a>0.$
\end{enumerate}
Then 
$$
U\stackrel{M}{\star}f(x)={MU}(\rho)[x^{\rho}l(x^{-1})](1+O(g(x^{-1}))+O(h(x^{-1})))
$$
as $x\rightarrow 0.$

\end{theorem}

\it Proof. \rm Theorem \ref{T:MC0} follows from Theorem \ref{T:Atg}, applied to the functions 
$\widetilde{f}(x)=f\left(x^{-1}\right)$ and $\widetilde{U}(x)=U\left(x^{-1}\right)$.
Here we take into account (\ref{E:mel}), (\ref{E:inv}), and (\ref{E:min}).

\section{The Heston model}\label{S:Hmod}
In this section, we gather several known results for the Heston model, which is a popular stochastic volatility model. It will be assumed in the sequel that the interest rate $r$ is equal to zero. The stock price process $X$ and the variance process $Y$ in the Heston model satisfy the following system of stochastic differential equations:
\begin{equation}
\left\{\begin{array}{ll}
dX_t=\mu X_tdt+\sqrt{Y_t}X_tdW_t \\
dY_t=(a-bY_t)dt+c\sqrt{Y_t}dZ_t,
\end{array}
\right.
\label{E:Heston1}
\end{equation}
where $\mu\in\mathbb{R}$, $a\ge 0$, $b\ge 0$, $c> 0$. In (\ref{E:Heston1}), $W$ and $Z$ are correlated standard Brownian motions such that $d\langle W,Z\rangle_t=\rho dt$ with $\rho\in(-1,1)$. In the Heston model, the distribution $\mu_t$ of the stock price $X_t$ admits density $D_t^{(1)}$. The initial conditions for the processes 
$X$ and $Y$ will be denoted by
$x_0$ and $y_0$, respectively. The Heston model was introduced and studied in \cite{H}.

We have
$$
X_t=x_0\exp\left\{\mu t-\frac{1}{2}\int_0^t Y_s ds+\int_0^t \sqrt{Y_s} dW_s\right\},
$$
and the following formulas hold for the density $D_t^{(1)}$ in the Heston model in the case where $\mu=0$ and $x_0=1$:
\begin{equation}
D_t^{(1)}(x)=A_{1}x^{-A_{3}}\exp\left\{A_{2}\sqrt{\log x}\right\}\left( \log
x\right) ^{-\frac{3}{4}+\frac{a}{c^{2}}}
\bigl(1+O((\log x)^{-\frac{1}{2}})\bigr)  \label{E:0}
\end{equation}
as $x\rightarrow \infty$, and
\begin{equation}
    D_t^{(1)}(x)=\widetilde{A}_1 x^{\widetilde{A}_3}\exp\left\{\widetilde{A}_2 \sqrt{\log\frac{1}{x}}\right\} 
    \left(\log\frac{1}{x}\right)^{-\frac{3}{4}+\frac{a}{c^2}}
     \left(1+O\left(\left(\log\frac{1}{x}\right)^{-\frac{1}{2}}\right)\right)
 \label{E:est}
\end{equation}
as $x\rightarrow 0$. The constants appearing in formulas (\ref{E:0}) and (\ref{E:est}) will be described below. Formulas (\ref{E:0}) and (\ref{E:est}) were obtained in \cite{GS2} in the case where $\rho=0$, 
and in \cite{FGGS} for $-1<\rho< 0$. A more detailed discussion of those and similar results can be found in \cite{G2}. 

For general $x_0$ and $\mu$, we obtain
\begin{equation}
D_t^{(1)}(x)=B_{1}x^{-A_{3}}\exp\left\{A_{2}\sqrt{\log x}\right\}\left( \log
x\right) ^{-\frac{3}{4}+\frac{a}{c^{2}}}
\bigl(1+O((\log x)^{-\frac{1}{2}})\bigr)  \label{E:0S}
\end{equation}
as $x\rightarrow \infty$, and
\begin{equation}
    D_t^{(1)}(x)=\widetilde{B}_1 x^{\widetilde{A}_3}\exp\left\{\widetilde{A}_2 \sqrt{\log\frac{1}{x}}\right\} 
    \left(\log\frac{1}{x}\right)^{-\frac{3}{4}+\frac{a}{c^2}}
     \left(1+O\left(\left(\log\frac{1}{x}\right)^{-\frac{1}{2}}\right)\right)
 \label{E:estS}
\end{equation}
as $x\rightarrow 0$. In (\ref{E:0S}) and (\ref{E:estS}), the constants $B_1$ and $\widetilde{B}_1$ are defined as follows:
$$
B_1=A_1\left(x_0e^{\mu t}\right)^{-A_3-1}
$$
and
$$
\widetilde{B}_1=\widetilde{A}_1\left(x_0e^{\mu t}\right)^{\widetilde{A}_3-1}.
$$
The proof of (\ref{E:estS}) uses (\ref{E:0}) and the following simple formulas:
$$
\exp\left\{r_1\sqrt{\log x+r_2}\right\}=\exp\left\{r_1\sqrt{\log x}\right\}
\bigl(1+O((\log x)^{-\frac{1}{2}})\bigr),\quad x\rightarrow\infty,\quad r_1\in\mathbb{R},\quad r_2\in\mathbb{R}, 
$$
and 
$$
(\log x+r_3)^{r_4}=(\log x)^{r_4}\bigl(1+O((\log x)^{-\frac{1}{2}})\bigr),\quad x\rightarrow\infty,\quad r_3\in\mathbb{R},\quad r_4\in\mathbb{R}.
$$
The proof of (\ref{E:estS}) is similar. It is based on formula (\ref{E:est}).

We will next provide explicit formulas for the constants appearing in formulas (\ref{E:0}) and (\ref{E:est}).
Given $s\ge 1$, define the explosion time for the moment of order~$s$ by
\[
T^{\ast }( s)=\sup \left\{
t\geq 0:\mathbb{E}[ X_{t}^s] <\infty \right\},
\]
and for any $t> 0$, let the $s_{+}$ be the upper critical moment defined by
$$
s_{+}=s_{+}(t)=\sup \left\{ s\geq1 : \mathbb{E}[X_{t}^{s}]<\infty\right\}.
$$
For the Heston model, the explosion time~$%
T^{\ast }$ is explicitly known (see \cite{AP,K-R}). The critical moment, for fixed~$t$, can then
be determined from $T^{\ast }( s_+(t)) =t$. The previous equality shows that
$s_{+}(t) \geq 1$ is the generalized inverse of the
function~$T^{\ast }(\cdot) $.

The lower critical moment is defined as follows:
$$
s_{-}=s_{-}(t)=\inf \left\{ s\leq0 : \mathbb{E}[X_t^{s}]<\infty \right\}.
$$
For fixed $t> 0$, the quantities
$$
\sigma_{+}=-\left. \frac{\partial T^{\ast }( s) }{\partial s}%
\right\vert _{s=s_+} \qquad \text{and} \qquad
\kappa_{+}= \left. \frac{\partial^2 T^{\ast }( s) }{\partial s^2}%
\right\vert _{s=s_+} 
$$
are called the upper critical slope and the upper critical curvature,
respectively.
Similarly, the lower critical slope and curvature are defined by
\begin{equation*}
\sigma_{-}=-\left. \frac{\partial T^{\ast }( s) }{\partial s}%
\right\vert _{s=s_{-}} \qquad \text{and} \qquad
\kappa_{-}= \left. \frac{\partial^2 T^{\ast }( s) }{\partial s^2}%
\right\vert _{s=s_-}
\end{equation*}
respectively.

In formula (\ref{E:0}), the constants $A_1$, $A_2$, and $A_3$ are given by
\begin{align*}
   & A_1 = \frac{1}{\sqrt{\pi }} 2^{-\frac{3}{4}-\frac{a}{c^{2}}}y_{0}^{\frac{1}{4}-\frac{a}{c^{2}}}
    c^{\frac{2a}{c^{2}}-\frac{1}{2}}\sigma_{+}^{-\frac{a}{c^{2}}-\frac{1}{4}}  \\
    & \quad \times \exp\left\{-y_0 \left(\frac{c\rho s_{+}-b}{c^2}+\frac{\kappa_{+}}{c^2\sigma_{+}^2}\right)
     -\frac{at}{c^2}(c\rho s_{+}-b)\right\}  \\
    & \quad \times
    \left\{\frac{2\sqrt{(b-c\rho s_{+})^2+c^2(s_{+}-s_{+}^2)}}
    {c^2s_{+}(s_{+}-1)\sinh\left[\frac{t}{2}\sqrt{(b-c\rho s_{+})^2+c^2(s_{+}-s_{+}^2)}\right]}\right\}^{\frac{2a}{c^2}},
\end{align*}
$$
A_{2}=2\sqrt{2y_{0}}c^{-1}\sigma_{+}^{-\frac{1}{2}},
$$
and
$$
  A_{3}=s_{+}+1.
$$
In addition, the constants $\widetilde{A}_1$, $\widetilde{A}_2$, and $\widetilde{A}_3$ in formula (\ref{E:est}) are as follows:

  \begin{align*}
    &\widetilde{A}_1 = \frac{1}{2\sqrt{\pi }} \left( 2y_{0}\right) ^{1/4-a/c^{2}}c^{2a/c^{2}-1/2}\sigma_{-}^{-a/c^{2}-1/4}  \\
      & \exp\left\{ -y_0 \left( \frac{s_{-} \rho c-b}{c^2}+\frac{\kappa_{-}}{c^2\sigma_{-}^2} \right)
      -\frac{at}{c^2}(c\rho s_{-}-b) \right\}  \\
      & 
      \left(\frac{2\sqrt{b^2-2bc\rho s_{-}+c^2 s_{-}(1-(1-\rho^2)s_{-})}}{c^2 s_{-}(s_{-}-1)
      \sinh \frac{t}{2}\sqrt{b^2-2bc\rho s_{-}+c^2 s_{-}(1-(1-\rho^2)s_{-})} } \right)^{\frac{2a}{c^2}},
     \end{align*}
 $$
  \widetilde{A}_{2} 
    =2\sqrt{2y_{0}}c^{-1}\sigma_{-}^{-\frac{1}{2}},
  $$
and
$$
\widetilde{A}_{3}=-(s_{-}+1).
$$
Note that the constants described above depend on $t$. 
\begin{remark}\label{R:const} \rm
It follows from (\ref{E:0}) and (\ref{E:est}) that the interval $(-A_3,\widetilde{A}_3)$ 
belongs to the domain of the Mellin transform $M D_t^{(1)}$.
\end{remark}

\section{The Heston model with double exponential jumps}\label{S:ku}
Let $N$ be a standard Poisson process with intensity $\lambda>0$, and consider a compound Poisson process defined by
$$
J_t=\sum_{i=1}^{N_t} (V_i-1),\quad t\ge 0,
$$
where $V_i$ are positive identically distributed random variables such that the distribution density $g$ of the
random variable $U_i=\log V_i$ is double exponential. This means that
$$
g(u)=p\eta_1 e^{-\eta_1 u}\chi_{\{u\geq 0\}}+q\eta_2 e^{\eta_2 u}\chi_{\{u<0\}}.
$$
where $\eta_1>1,$ $\eta_2>0,$ and $p$ and $q$ are positive numbers such that $p+q=1.$
The condition $\eta_1>1$ is necessary and sufficient for the random variable $J_t$ to have finite expectation. 

S. Kou introduced and studied a perturbation of the Black-Scholes model based on the jump process described above 
(see \cite{K}, see also \cite{KW}). In \cite{GV}, we considered a similar perturbation of the Heston model. The stock price process and the variance process $Y$ in the perturbed Heston model
satisfy the following system of stochastic differential equations:
\begin{equation}
\left\{\begin{array}{ll}
d{\widetilde X}_t=\mu {\widetilde X}_{t-}dt+\sqrt{Y_t}{\widetilde X}_{t-}dW_t+{\widetilde X}_{t-}dJ_t\\
dY_t=\left(a-bY_t\right)dt+c\sqrt{Y_t}dZ_t.
\end{array}
\right.
\label{E:H1}
\end{equation}
It is assumed in (\ref{E:H1}) that the compound Poisson process $J$ is independent of standard 
Brownian motions $W$ and $Z$. The initial conditions for the processes $\widetilde{X}$ and $Y$ will be 
denoted by $x_0$ and $y_0$, respectively.

It is not hard to see that
\begin{equation}
{\widetilde X}_t=x_0\exp\left\{\mu t-\frac{1}{2}\int_0^t Y_s ds+\int_0^t \sqrt{Y_s} dW_s
+\sum_{i=1}^{N_t} U_i \right\}.
\label{D2}
\end{equation}
The validity of the equality in (\ref{D2}) follows from the Dol\'{e}ans-Dade formula (see, for example, \cite{P}).

The Heston model with double exponential jumps is a mixed stochastic model. Indeed, using formula (\ref{D2}), we can split the process $\widetilde{X}$ into the product of the following processes:
\begin{equation}
X^{(1)}_t=x_0\exp\left\{\mu t-\frac{1}{2}\int_0^t Y_s ds+\int_0^t \sqrt{Y_s} dW_s\right\}
\label{E:razd1}
\end{equation}
and 
\begin{equation}
X^{(2)}_t=\exp\left\{\sum_{i=1}^{N_t} U_i \right\}.
\label{E:razd2}
\end{equation}
Note that $X^{(1)}=X$, where $X$ is the stock price process in the 
Heston model described by (\ref{E:Heston1}). Note also that for every $t\ge 0$, we have
$\mathbb{E}\left[X^{(2)}_t\right]<\infty$ (see Remark \ref{R:integ} below).

Let us put 
$T_t=\sum_{i=1}^{N_t} U_i$. The distribution $\mu_t$ of the random variable $T_t$ is given by
\begin{align}
d\mu_t(y)&=e^{-\lambda t}d\delta_0(y) 
+\left[G_1(t,y)e^{-\eta_1 y}\chi_{\{y> 0\}}+G_2(t,y)e^{\eta_2 y}
\chi_{\{y< 0\}}\right]dy,
\label{E:drv}
\end{align}
where $y\in(-\infty,\infty)$ and $\delta_0$ is the delta-measure at $y=0$. The functions $G_1$ and $G_2$ in
the previous formula are defined by
\begin{equation}
G_1(t,u)=\sum_{k=0}^{\infty}a_ku^k,\quad u> 0,
\label{E:labb1}
\end{equation}
with
\begin{equation}
a_k=\frac{\eta_1^{k+1}}{k!}\sum_{n=k+1}^{\infty}\pi_nP_{n,k+1},
\label{E:with1}
\end{equation}
and
$$
G_2(t,-u)=\sum_{k=0}^{\infty}b_ku^k,\quad u> 0,
$$
with
\begin{equation}
b_k=\frac{\eta_2^{k+1}}{k!}\sum_{n=k+1}^{\infty}\pi_nQ_{n,k+1}.
\label{E:with2}
\end{equation}
The numbers $\pi_n$ in the previous formulas depend on $t$. They are defined by
$\pi_0=e^{-\lambda t}$ and $\pi_n=e^{-\lambda t}(\lambda t)^n(n!)^{-1}$ for all $n\ge 1$. In addition,
the numbers $P_{n,k}$ and $Q_{n,k}$ are given by
\begin{equation}
P_{n,k}=\sum_{i=k}^{n-1}\left(\begin{array}{c}
      n-k-1 \\
     i-k
    \end{array}
  \right) 
\left(
    \begin{array}{c}
      n \\
      i
    \end{array}
  \right) 
\left(\frac{\eta_1}{\eta_1+\eta_2}\right)^{i-k}\left(\frac{\eta_2}{\eta_1+\eta_2}\right)^{n-i}p^iq^{n-i}
\label{E:with3}
\end{equation}
for all $1\le k\le n-1$, and
$$
Q_{n,k}=\sum_{i=k}^{n-1}\left(\begin{array}{c}
      n-k-1 \\
     i-k
    \end{array}
  \right) 
\left(
    \begin{array}{c}
      n \\
      i
    \end{array}
  \right) 
\left(\frac{\eta_1}{\eta_1+\eta_2}\right)^{n-i}\left(\frac{\eta_2}{\eta_1+\eta_2}\right)^{i-k}p^{n-i}q^i
$$
for all $1\le k\le n-1$. We also have $P_{n,n}=p^n$ and $Q_{n,n}=q^n$. Formula (\ref{E:drv}) can be derived using
Proposition B.1 in \cite{K} (see the derivation in \cite{GV}, or in Section 10.8 of \cite{G2}).

It follows from (\ref{E:drv}) that the distribution $\mu^{(2)}_t$ of the random variable
$X^{(2)}_t$ satisfies
\begin{equation}
d\mu^{(2)}_t(x)=e^{-\lambda t}d\delta_1(x)+H(t,x)dx,\quad x> 0.
\label{E:drvi}
\end{equation}
In (\ref{E:drvi}), $\delta_1$ is the delta-measure at $x=1$, the function $H$ is defined by
\begin{equation}
H(t,x)=H_1(t,x)x^{-\eta_1-1}\chi_{\{x> 1\}}+H_2(t,x)x^{\eta_2-1}\chi_{\{0< x< 1\}},
\label{E:HH}
\end{equation}
where 
\begin{equation}
H_1(t,x)=G_1(t,\log x),\quad x> 1,
\label{E:Hc1}
\end{equation} 
and 
\begin{equation}
H_2(t,x)=G_2(t,\log x),\quad 0< x< 1.
\label{E:Hc2} 
\end{equation}

The next assertion provides useful approximations to the coefficients $a_k$ and $b_k$ appearing in (\ref{E:with1}) and (\ref{E:with2}). 
\begin{theorem}\label{T:asapr}
There exist positive constants $c_1$ and $c_2$, independent of $k$, and such that 
\begin{equation}
0<a_k-\widehat{a}_k\le c_1\frac{\widehat{a}_k}{k+1},\quad k\ge 0,
\label{E:o3}
\end{equation}
and
\begin{equation}
0< b_k-\widehat{b}_k\le c_2\frac{\widehat{b}_k}{k+1},\quad k\ge 0,
\label{E:o4}
\end{equation}
where
$$
\widehat{a}_k=\exp\left\{\frac{\eta_2\lambda tq}{\eta_1+\eta_2}-\lambda t\right\}\frac{\left(\eta_1\lambda tp\right)^{k+1}}{k!(k+1)!}
,\quad k\ge 0,
$$
and
$$
\widehat{b}_k=\exp\left\{\frac{\eta_1\lambda tp}{\eta_1+\eta_2}-\lambda t\right\}\frac{\left(\eta_2\lambda tq\right)^{k+1}}{k!(k+1)!}
,\quad k\ge 0.
$$
\end{theorem}

\it Proof. \rm For $i\ge 1$ and $m\ge i+1$, put
$$
\gamma_{k+m,k+i}=\left(\begin{array}{c}
      m-2 \\
     i-1
    \end{array}
  \right) 
  \left(\begin{array}{c}
      k+m \\
     k+i
    \end{array}
  \right)
  \left(\frac{\eta_1}{\eta_1+\eta_2}\right)^{i-1}\left(\frac{\eta_2}{\eta_1+\eta_2}\right)^{m-i}
  p^{k+i}q^{m-i}. 
$$

It follows from (\ref{E:with1}) and (\ref{E:with3}) that for all $k\ge 1$,
\begin{equation}
a_k=\sum_{i=0}^{\infty}a_{k,i},
\label{E:o5}
\end{equation}
where
$$
a_{k,0}=\frac{\eta_1^{k+1}}{k!}\pi_{k+1}p^{k+1}
$$
and
\begin{equation}
a_{k,i}=\frac{\eta_1^{k+1}}{k!}\sum_{m=i+1}^{\infty}\pi_{k+m}\gamma_{k+m,k+i}
\label{E:o7}
\end{equation}
for all $i\ge 1$. We have
\begin{align}
a_{k,0}+a_{k,1}&=e^{-\lambda t}\frac{(\eta_1\lambda tp)^{k+1}}{k!(k+1)!}\left[1+\sum_{m=2}^{\infty}\frac{1}{(m-1)!}
\left(\frac{\eta_2\lambda tq}{\eta_1+\eta_2}\right)^{m-1}\right] \nonumber \\
&=\exp\left\{\frac{\eta_2\lambda tq}{\eta_1+\eta_2}-\lambda t\right\}\frac{\left(\eta_1\lambda tp\right)^{k+1}}{k!(k+1)!}
=\widehat{a}_k.
\label{E:o8}
\end{align}
Therefore, (\ref{E:o5}), (\ref{E:o7}), and (\ref{E:o8}) imply the following:
\begin{align}
&0\le a_k-\widehat{a}_k=\sum_{i=2}^{\infty}a_{k,i} \nonumber \\
&=e^{-\lambda t}
\frac{\left(\eta_1\lambda tp\right)^{k+1}}{k!(k+1)!}\sum_{i=2}^{\infty}\frac{1}{(k+i)\cdots(k+2)(i-1)!}
\left(\frac{\eta_1}{\eta_1+\eta_2}\right)^{i-1}p^{i-1} \nonumber \\
&\quad\sum_{m=i+1}^{\infty}\frac{(\lambda t)^{m-1}(m-2)!}{(m-i-1)!(m-i)!}\left(\frac{\eta_2}
{\eta_1+\eta_2}\right)^{m-i}q^{m-i} \nonumber \\
&\le \beta_1\frac{\widehat{a}_k}{k+1}\sum_{i=2}^{\infty}\frac{1}{i!(i-1)!}
\left(\frac{\eta_1}{\eta_1+\eta_2}\right)^{i-1}(\lambda tp)^{i-1}\sum_{j=0}^{\infty}\frac{(\lambda t)^{j+1}(j+i-1)!}{j!(j+1)!}\left(\frac{\eta_2}
{\eta_1+\eta_2}\right)^{j+1}q^{j+1},
\label{E:o9}
\end{align}
where $\beta_1$ is a positive constant. For $i\ge 2$ and $j\ge 1$, we have
$$
\frac{(j+i-1)!}{i!j!}\le\frac{(j+i)^i}{i!}\le\beta_2e^i\left(1+\frac{j}{i}\right)^i\le\beta_2e^{i+j},
$$
where $\beta_2$ is a positive constant. In the proof of the previous estimates, we used Striling's formula. Now, it is not hard to see that
the last double series in (\ref{E:o9}) converges, and it follows from (\ref{E:o9}) that the estimate in
(\ref{E:o3}) is valid. The proof of the estimate in (\ref{E:o4}) is similar.

This completes the proof of Theorem \ref{T:asapr}.

We will next further simplify formulas (\ref{E:o3}) and (\ref{E:o4}). Set
$$
C_1=\frac{\eta_1\lambda tp}{2\pi}\exp\left\{\frac{\eta_2\lambda tq}{\eta_1+\eta_2}-\lambda t\right\},\quad
B_1=\eta_1\lambda tp,
$$
$$
C_2=\frac{\eta_2\lambda tq}{2\pi}\exp\left\{\frac{\eta_1\lambda tp}{\eta_1+\eta_2}-\lambda t\right\},\quad
B_2=\eta_2\lambda tq,
$$
and consider the following sequences:
\begin{equation}
d_0=C_1,\quad d_k=C_1\frac{B_1^ke^{2k}}{k^{2k+2}},\quad k\ge 1,
\label{E:d1}
\end{equation}
and 
$$
l_0=C_2,\quad l_k=C_2\frac{B_2^ke^{2k}}{k^{2k+2}},\quad k\ge 1.
$$
\begin{corollary}\label{C:cin}
The following formula holds:
\begin{equation}
|a_k-d_k|\le\frac{\alpha_1}{k+1}d_k,\quad k\ge 0,
\label{E:cin1}
\end{equation}
and
\begin{equation}
|b_k-l_k|\le\frac{\alpha_2}{k+1}l_k,\quad k\ge 0,
\label{E:cin4}
\end{equation}
where $\alpha_1$ and $\alpha_2$ are some positive constants.
\end{corollary}

\it Proof. \rm We will need Stirling's formula in the asymptotic form:
\begin{equation}
n!=\sqrt{2\pi}n^{n+\frac{1}{2}}e^{-n}\left(1+O\left(\frac{1}{n}\right)\right)
\label{E:Sf}
\end{equation}
as $n\rightarrow\infty$. It is not hard to see, using (\ref{E:o3}) and (\ref{E:Sf}) that
\begin{align*}
a_k&=\exp\left\{\frac{\eta_2\lambda tq}{\eta_1+\eta_2}-\lambda t\right\}\frac{\left(\eta_1\lambda tp\right)^{k+1}}
{k!k!k\left(1+\frac{1}{k}\right)}
\left(1+O\left(\frac{1}{k+1}\right)\right) \\
&=\exp\left\{\frac{\eta_2\lambda tq}{\eta_1+\eta_2}-\lambda t\right\}\frac{\left(\eta_1\lambda tp\right)^{k+1}}
{k!k!k}
\left(1+O\left(\frac{1}{k+1}\right)\right) \\
&=\frac{1}{2\pi}\exp\left\{\frac{\eta_2\lambda tq}{\eta_1+\eta_2}-\lambda t\right\}
\frac{\left(\eta_1\lambda tp\right)^{k+1}e^{2k}}
{k^{2k+2}}
\left(1+O\left(\frac{1}{k+1}\right)\right)
\end{align*}
as $k\rightarrow\infty$. This establishes (\ref{E:cin1}). The proof of (\ref{E:cin4}) is similar.
\subsection{Properties of the functions $H_1$ and $H_2$}\label{SS:H1H2}
In the present subsection, we study the asymptotic behavior of the functions $H_1$ and $H_2$ defined in (\ref{E:Hc1}) and (\ref{E:Hc2}). It will be shown first that these functions are of slow variation with remainder. 
\begin{lemma}\label{L:ksv}

For every $t> 0$, the functions $x\mapsto H_1(t,x)$ and $x\mapsto H_2(t,x^{-1})$ belong to the Zygmund class ${\cal Z}$.

\end{lemma}

\it Proof. \rm
Let us fix $t> 0$. Since the function $x\mapsto H_1(t,x)$ is increasing on $x>1,$ 
the function $\phi_{\alpha}(t,x)=x^{\alpha}H_1(t,x)$, where $\alpha> 0$, is also increasing. It remains to prove 
that the function $\psi_{\alpha}(t,x)=x^{-\alpha}H_1(t,x)$ is ultimately decreasing. We have 

$$\psi^{\prime}_{\alpha}(t,x)=-\alpha x^{-\alpha-1} G_1(t,\log x)+x^{-\alpha-1}G^{\prime}_{1}(t,\log x).$$ 
Therefore, the condition $\psi^{\prime}_{\alpha}(t,x)\le 0$ is equivalent to the condition 

$$\frac{G^{\prime}_{1}(t,\log x)}{G_1(t,\log x)}\leq \alpha$$
for all $x>x_{\alpha},$ which in its turn is equivalent to the condition 

\begin{equation}\label{E:limit1}
\frac{G^{\prime}_{1}(t,\log x)}{G_1(t,\log x)}\rightarrow 0
\end{equation}
as $x\rightarrow \infty.$ Now it is clear that it suffices to prove (\ref{E:limit1}).

Using the definition of the function $G_1$, we obtain

\begin{equation}\label{limit2}
\frac{G^{\prime}_{1}(t,\log x)}{G_1(t,\log x)}=\frac{\sum_{k=1}^{\infty} a_k k (\log x)^{k-1}}
{\sum_{k=0}^{\infty} a_k (\log x)^{k}}
\end{equation}
where the coefficients $a_k$ are defined by (\ref{E:with1}).
It is not hard to see using (\ref{E:o3}) that for all $k\ge 2$,
\begin{equation}
\frac{ka_k}{a_{k-1}}\le\frac{c}{k+1}
\label{E:som1}
\end{equation}
with some $c> 0$. Hence for every $\varepsilon>0$ there exists a positive integer 
$k_{\varepsilon}$ such that $ka_k\leq \varepsilon a_{k-1}$
for all $k>k_{\varepsilon}.$ It follows from (\ref{limit2}) that  

\begin{equation}\label{limit5}
\frac{G^{\prime}_{1}(t,\log x)}{G_1(t,\log x)}\leq \frac{\sum_{k=1}^{k_{\varepsilon}} a_k k (\log x)^{k-1}}
{\sum_{k=1}^{\infty} a_k (\log x)^{k}}+\varepsilon.
\end{equation}
It is clear that for fixed $\varepsilon$ the first term on the right-hand side of (\ref{limit5}) tends to $0$ as $x\rightarrow \infty.$ 
Now, it is not hard to see that condition (\ref{E:limit1}) holds.

This completes the proof of Lemma \ref{L:ksv} for the function 
$x\mapsto H_1(t,x)$. The proof for the function $x\mapsto H_2(t,\frac{1}{x})$ is similar. 
\begin{remark}\label{R:integ} \rm
It follows from (\ref{E:drvi}), the fact that ${\cal Z}\subset R_0$, and Lemma \ref{L:ksv} that
the process $t\mapsto X^{(2)}_t$ is an integrable process.
\end{remark}
\begin{lemma}\label{L:ksv2}
For every $t> 0$, the functions 
$x\mapsto H_1(t,x)$ and $x\mapsto H_2(t,x^{-1})$
belong to the class $R_0^g$, where the function $g$ is given by
$g(x)=(\log x)^{-\frac{1}{2}}$.
\end{lemma}

\it Proof. \rm The function $x\mapsto H_1(t,x)$ is an increasing differentiable function from the Zygmund class.
Therefore, it suffices to prove that there exists $\widetilde{c}> 0$ such that
\begin{equation}
\frac{xH_1^{\prime}(t,x)}{H_1(t,x)}\le\widetilde{c}(\log x)^{-\frac{1}{2}}
\label{E:eqx1}
\end{equation}
for all $x> x_0$ (see Corollary \ref{C:use} and (\ref{E:forn})). 
It is easy to see that the estimate in (\ref{E:eqx1})
is equivalent to the following:
\begin{equation}
\frac{G_1^{\prime}(t,\log x)}{G_1(t,\log x)}=O\left((\log x)^{-\frac{1}{2}}\right)
\label{E:eqx2}
\end{equation}
as $x\rightarrow\infty$.

We have 
\begin{align}
&\frac{G_1^{\prime}(t,\log x)}{G_1(t,\log x)}=(\log x)^{-1}\frac{\sum_{k=1}^{\infty}a_kk(\log x)^k}
{\sum_{k=0}^{\infty}a_k(\log x)^k}
\nonumber 
\\
&=(\log x)^{-1}\frac{\sum_{k=1}^{[\sqrt{\log x}]}a_kk(\log x)^k
+\sum_{k=[\sqrt{\log x}]+1}^{\infty}a_kk(\log x)^k}{\sum_{k=0}^{\infty}a_k(\log x)^k}
\nonumber \\
&\le(\log x)^{-\frac{1}{2}}+(\log x)^{-1}\frac{\sum_{k=[\sqrt{\log x}]+1}^{\infty}a_kk(\log x)^k}{\sum_{k=0}^{\infty}a_k(\log x)^k}.
\label{E:eqx3}
\end{align}
Next, using (\ref{E:som1}) in (\ref{E:eqx3}), we obtain
\begin{align*}
&\frac{G_1^{\prime}(t,\log x)}{G_1(t,\log x)}\le(\log x)^{-\frac{1}{2}}
+(\log x)^{-1}\frac{\sum_{k=[\sqrt{\log x}]+1}^{\infty}\frac{ca_{k-1}}{k+1}(\log x)^k}
{\sum_{k=1}^{\infty}a_k(\log x)^k}
\\
&\le(\log x)^{-\frac{1}{2}}+c(\log x)^{-\frac{3}{2}}
\frac{\sum_{k=[\sqrt{\log x}]+1}^{\infty}a_{k-1}(\log x)^k}{\sum_{k=1}^{\infty}a_k(\log x)^k} 
\\
&=O\left((\log x)^{-\frac{1}{2}}\right)
\end{align*}
as $x\rightarrow\infty$. This establishes estimate (\ref{E:eqx2}).

The proof of Lemma \ref{L:ksv2} in the case of the function $x\mapsto H_1(t,x)$ is thus completed.
For the function $x\mapsto H_2\left(t,\frac{1}{x}\right)$, the proof is similar.

Recall that $G_1(t,\cdot)=\sum_{k=0}^{\infty}a_ku^k$ (see (\ref{E:labb1})), where the coefficients $a_k$ are given by (\ref{E:with1}). Define two auxiliary functions $\widetilde{G}_1(t,\cdot)$ and $\widehat{G}_1(t,\cdot)$
as follows:
$$
\widetilde{G}_1(t,u)=\sum_{k=0}d_ku^k\quad\mbox{and}\quad \widehat{G}_1(t,u)=\sum_{k=0}^{\infty}\frac{d_k}{k+1}u^k,
$$
where $u\ge 0$ and the sequence $d$ is given by (\ref{E:d1}). Then (\ref{E:cin1}) implies that
$$
|G_1(t,u)-\widetilde{G}_1(t,u)|\le\alpha_1\widehat{G}_1(t,u).
$$

The functions $\widetilde{G}_1$ and $\widehat{G}_1$ are defined as sums of certain power series. Our next goal is to compare these functions with some standard functions. Analyzing the structure of the coefficients $d_k$, we guess that the following family of functions may be useful:
\begin{equation}
\lambda_{s,r}(u)=s\cosh(r\sqrt{u})=\sum_{k=0}^{\infty}\widetilde{d}_ku^k,\quad u\ge 0,\quad r> 0,\quad s> 0,
\label{E:nz1}
\end{equation}
where
$$
\widetilde{d}_k=s\frac{r^{2k}}{(2k)!}.
$$
It is clear that $\widetilde{d}_0=s$. Moreover, using Stirling's formula, we see that
$$
\widetilde{d}_k=s\frac{r^{2k}e^{2k}}{2\sqrt{\pi}2^{2k}k^{2k+\frac{1}{2}}}
\left(1+O\left(\frac{1}{k}\right)\right)
$$
as $k\rightarrow\infty$. Next, comparing the coefficients $d_k$ and $\widetilde{d}_k$, we see that if we set
\begin{equation}
s=2\sqrt{\pi}C_1\quad\mbox{and}\quad r=2\sqrt{B_1},
\label{E:axit}
\end{equation}
then 
\begin{equation}
|d_k-(k+1)^{-\frac{3}{2}}\widetilde{d}_k|\le\delta_1(k+1)^{-\frac{5}{2}}\widetilde{d}_k
\label{E:axit1}
\end{equation}
for some $\delta_1> 0$ and all $k\ge 0$. Finally, it follows from (\ref{E:cin1}) and (\ref{E:axit1}) that the coefficients
$a_k$ and $\widetilde{d}_k$ satisfy the following condition:
\begin{equation}
|a_k-(k+1)^{-\frac{3}{2}}\widetilde{d}_k|\le\delta_2(k+1)^{-\frac{5}{2}}\widetilde{d}_k
\label{E:axit2}
\end{equation}
for some $\delta_2> 0$ and all $k\ge 0$.
\subsection{The Riemann-Liouville integrals}\label{S:RLI}
In this subsection, we consider only fractional integrals of functions given by everywhere convergent power series with positive coefficients. Let $f(u)=\sum_{n=0}^{\infty}c_nu^n$ be a function on $\mathbb{R}$ such that $c_n> 0$ for all $n\ge 0$ and the function $F(z)=\sum_{n=0}^{\infty}c_nz^n$ is an entire function on $\mathbb{C}$. For $\alpha< 0$, 
the Riemann-Liouville fractional integral $D^{\alpha}f$ is defined as follows:
\begin{equation}
D^{\alpha}f(u)=\frac{1}{\Gamma(-\alpha)}\int_0^{u}f(y)(u-y)^{-\alpha-1}dy.
\label{E:rl}
\end{equation}
Then the following formula is valid:
\begin{equation}
D^{\alpha}f(u)=u^{-\alpha}\sum_{n=0}^{\infty}c_{n,\alpha}c_nu^n,
\label{E:h1}
\end{equation}
where
$$
c_{n,\alpha}=\frac{\Gamma(n+1)}{\Gamma(n-\alpha+1)}
$$
(see the definition of the Riemann-Liouville integral in \cite{LOT}, Section 5, and formula (3.2) in \cite{LOT}, Section 3). We will next charactize the asymptotic behavior of the sequence $c_{n,\alpha}$.
Using the asymptotic formula for the Gamma function, that is, the formula
$$
\Gamma(u)=\sqrt{2\pi}u^{u-\frac{1}{2}}e^{-u}\left(1+O\left(\frac{1}{u}\right)\right)
$$
as $u\rightarrow\infty$, we obtain
\begin{align*}
c_{n,\alpha}&=\frac{e^{|\alpha|}(n+1)^{n+|\alpha|+\frac{1}{2}}}
{(n+|\alpha|+1)^{n+|\alpha|+\frac{1}{2}}(n+1)^{|\alpha|}}
\left(1+O\left(\frac{1}{n}\right)\right) \\
&=e^{|\alpha|}\left(\frac{n+1}{n+|\alpha|+1}\right)^{n+|\alpha|+1}
\frac{(n+|\alpha|+1)^{\frac{1}{2}}}{(n+1)^{\frac{1}{2}}}\frac{1}{(n+1)^{|\alpha|}}
\left(1+O\left(\frac{1}{n}\right)\right) \\
&=e^{|\alpha|}\left(1-\frac{|\alpha|}{n+|\alpha|+1}\right)^{n+|\alpha|+1}
\frac{1}{(n+1)^{|\alpha|}}
\left(1+O\left(\frac{1}{n}\right)\right)
\end{align*}
as $n\rightarrow\infty$.

It is not hard to prove that for every $c> 0$, 
$$
\left(1-\frac{c}{x}\right)^x=e^{-c}\left(1+O\left(\frac{1}{x}\right)\right)
$$
as $x\rightarrow\infty$. Therefore,
$$
c_{n,\alpha}=\frac{1}{(n+1)^{|\alpha|}}\left(1+O\left(\frac{1}{n+1}\right)\right)
$$
as $n\rightarrow\infty$, and hence
\begin{equation}
\frac{1}{(n+1)^{|\alpha|}}=c_{n,\alpha}\left(1+O\left(\frac{1}{n+1}\right)\right)
\label{E:ms}
\end{equation}
as $n\rightarrow\infty$. Since
$$
\frac{c_{n,\alpha}}{n+1}=\frac{\Gamma(n+1)}{(n+1)\Gamma(n+|\alpha|+1)}
\le c\frac{\Gamma(n+1)}{(n+|\alpha|+1)\Gamma(n+|\alpha|+1)}=
c\frac{\Gamma(n+1)}{\Gamma(n+|\alpha|+2)}=c\cdot c_{n,\alpha-1},
$$
formula (\ref{E:ms}) implies that there exists a constant $\delta_3> 0$ for which
\begin{equation}
\left|\frac{1}{(n+1)^{|\alpha|}}-c_{n,\alpha}\right|\le\delta_3c_{n,\alpha-1}
\label{E:mss}
\end{equation}
for all $n\ge 0$.

Our next goal is to combine various estimates for the coefficients obtained above.
\begin{lemma}\label{L:combi}
There exists a constant $\delta_4> 0$ such that
$$
\left|a_k-c_{k,-\frac{3}{2}}\widetilde{d}_k\right|\le\delta_4c_{k,-\frac{5}{2}}\widetilde{d}_k
$$
for all $k\ge 0$.
\end{lemma}

\it Proof. \rm It follows from (\ref{E:axit2}), (\ref{E:ms}), and (\ref{E:mss}) that
\begin{align*}
&\left|a_k-c_{k,-\frac{3}{2}}\widetilde{d}_k\right|\le\left|a_k-(k+1)^{-\frac{3}{2}}\widetilde{d}_k\right|
+\left|c_{k,-\frac{3}{2}}-(k+1)^{-\frac{3}{2}}\right|\widetilde{d}_k \\
&\le\delta_2(k+1)^{-\frac{5}{2}}\widetilde{d}_k+\delta_3c_{k,-\frac{5}{2}}\widetilde{d}_k\le\delta_4c_{k,-\frac{5}{2}}\widetilde{d}_k.
\end{align*}

This completes the proof of Lemma \ref{L:combi}.

The following assertion can be easily derived from (\ref{E:labb1}), (\ref{E:nz1}), and (\ref{E:h1}).
\begin{theorem}\label{T:bles}
There exists a positive constant $c$ such that
\begin{equation}
\left|G_1(t,u)-u^{-\frac{3}{2}}D^{-\frac{3}{2}}\lambda_{s,r}(u)\right|
\le cu^{-\frac{5}{2}}D^{-\frac{5}{2}}\lambda_{s,r}(u)
\label{E:bles1}
\end{equation}
for all $u> 0$. In (\ref{E:bles1}), the values of the parameters $r$ and $s$ are chosen according to (\ref{E:axit}).
\end{theorem}

It follows from Theorem \ref{T:bles} that in order to understand the asymptotic behavior of the function $G_1(t,u)$ as 
$u\rightarrow\infty$, we have to study how the fractional integrals appearing in (\ref{E:bles1}) behave for large values of $u$. 
Using (\ref{E:nz1}) and (\ref{E:rl}), we obtain
\begin{align}
&u^{-\frac{3}{2}}D^{-\frac{3}{2}}\lambda_{s,r}(u)=\frac{s}{\Gamma\left(\frac{3}{2}\right)}
\int_0^1\cosh(r\sqrt{yu})(1-y)^{\frac{1}{2}}dy
\nonumber \\
&=\frac{s}{\Gamma\left(\frac{3}{2}\right)}\int_0^1e^{r\sqrt{u}z}z(1-z^2)^{\frac{1}{2}}dz
+\frac{s}{\Gamma\left(\frac{3}{2}\right)}\int_0^1e^{-r\sqrt{u}z}z(1-z^2)^{\frac{1}{2}}dz
\label{E:bles2}
\end{align}
and
\begin{align}
&u^{-\frac{5}{2}}D^{-\frac{5}{2}}\lambda_{s,r}(u)=\frac{s}{\Gamma\left(\frac{5}{2}\right)}
\int_0^1\cosh(r\sqrt{yu})(1-y)^{\frac{3}{2}}dy
\nonumber \\
&=\frac{s}{\Gamma\left(\frac{5}{2}\right)}\int_0^1e^{r\sqrt{u}z}z(1-z^2)^{\frac{3}{2}}dz
+\frac{s}{\Gamma\left(\frac{5}{2}\right)}\int_0^1e^{-r\sqrt{u}z}z(1-z^2)^{\frac{3}{2}}dz.
\label{E:bles3}
\end{align}

We have 
\begin{equation}
\int_0^1e^{-r\sqrt{u}z}z(1-z^2)^{\frac{1}{2}}dz=\int_0^{\frac{1}{2}}+\int_{\frac{1}{2}}^1
=O\left(u^{-1}\right).
\label{E:as1}
\end{equation}
Indeed, the third integral in (\ref{E:as1}) is $O\left(e^{-\frac{1}{2}r\sqrt{u}}\right)$, while the second integral can be estimated, using the integration by parts twice. Similarly,
\begin{equation}
\int_0^1e^{-r\sqrt{u}z}z(1-z^2)^{\frac{3}{2}}dz=
O\left(u^{-1}\right)
\label{E:as2}
\end{equation}
as $u\rightarrow\infty$. In addition,
$$
\int_0^1e^{r\sqrt{u}z}z(1-z^2)^{\frac{1}{2}}dz=e^{r\sqrt{u}}
\int_0^1e^{r\sqrt{u}(z-1)}z(1-z^2)^{\frac{1}{2}}dz=e^{r\sqrt{u}}
\int_0^1e^{-r\sqrt{u}y}(1-y)(2-y)^{\frac{1}{2}}y^{\frac{1}{2}}dy.
$$
For small values of $y$, we have $(1-y)(2-y)^{\frac{1}{2}}y^{\frac{1}{2}}=\sqrt{2}y^{\frac{1}{2}}
+c_1y^{\frac{3}{2}}+\cdots$. Using Watson's lemma (see \cite{BH}, p. 103), we obtain
\begin{equation}
\int_0^1e^{r\sqrt{u}z}z(1-z^2)^{\frac{1}{2}}dz=e^{r\sqrt{u}}
\left[\sqrt{2}\Gamma\left(\frac{3}{2}\right)r^{-\frac{3}{2}}u^{-\frac{3}{4}}+O\left(u^{-\frac{5}{4}}\right)\right]
\label{E:oo1}
\end{equation}
as $y\rightarrow\infty$. Similarly,
\begin{equation}
\int_0^1e^{r\sqrt{u}z}z(1-z^2)^{\frac{3}{2}}dz=e^{r\sqrt{u}}
\left[2^{\frac{3}{2}}\Gamma\left(\frac{5}{2}\right)r^{-\frac{5}{2}}u^{-\frac{5}{4}}+O\left(u^{-\frac{7}{4}}\right)\right]
\label{E:oo2}
\end{equation}
as $y\rightarrow\infty$.

Now we are ready to formulate and prove one of the main results of the present paper.
\begin{theorem}\label{T:mainre}
Let $H_1(t,\cdot)$ and $H_2(t,\cdot)$ be the functions defined by (\ref{E:Hc1}) and (\ref{E:Hc2}), respectively. 
Then the following asymptotic formulas hold:
\begin{align}
H_1(t,x)&=\frac{1}{2\sqrt{\pi}}(\eta_1\lambda tp)^{\frac{1}{4}}\exp\left\{\frac{\eta_2\lambda tq}
{\eta_1+\eta_2}-\lambda t\right\}(\log x)^{-\frac{3}{4}}
\nonumber \\
&\quad\exp\{2\sqrt{\eta_1\lambda tp}\sqrt{\log x}\}\left(1+O\left((\log x)^{-\frac{1}{2}}\right)\right)
\label{E:ur1}
\end{align}
as $x\rightarrow\infty$, and
\begin{align}
H_2(t,x)&=\frac{1}{2\sqrt{\pi}}(\eta_2\lambda tq)^{\frac{1}{4}}
\exp\left\{\frac{\eta_1\lambda tp}{\eta_1+\eta_2}-\lambda t\right\}\left(\log\frac{1}{x}\right)^{-\frac{3}{4}}
\nonumber \\
&\quad\exp\left\{2\sqrt{\eta_2\lambda tq}\sqrt{\log\frac{1}{x}}\right\}\left(1+O\left(\left(\log\frac{1}{x}\right)^{-\frac{1}{2}}\right)\right)
\label{E:ur2}
\end{align}
as $x\rightarrow 0$.
\end{theorem}

\it Proof. \rm Formula (\ref{E:ur1}) follows from (\ref{E:bles1}) and formulas (\ref{E:bles2}) - (\ref{E:oo2}). Here we take into account that the values of the parameters $r$ and $s$ are given by (\ref{E:axit}). The proof of formula (\ref{E:ur2}) uses the same ideas, and we leave it as an exercise for the interested reader.

\section{Asymptotic behavior of stock price densities in the Heston model with double exponential jumps}
\label{S:deg}
In this section, we study the asymptotic behavior of the marginal densities in the perturbed Heston model.
Our first goal is to characterize the asymptotics 
of the density of the absolutely continuous part $H$ of the distribution $\mu^{(2)}_t$ of the random variable
$X^{(2)}_t$. Recall that
$$
d\mu^{(2)}_t(x)=e^{-\lambda t}d\delta_1(x)+H(t,x)dx,\quad x> 0,
$$
where
$$
H(t,x)=H_1(t,x)x^{-\eta_1-1}\chi_{\{x> 1\}}+H_2(t,x)x^{\eta_2-1}\chi_{\{0< x< 1\}}
$$
(see (\ref{E:drvi})). It is clear from Theorem \ref{T:mainre} that the following assertion holds.
\begin{corollary}\label{C:amfi}
For every $t> 0$,
\begin{align}
H(t,x)&=\frac{1}{2\sqrt{\pi}}(\eta_1\lambda tp)^{\frac{1}{4}}\exp\left\{\frac{\eta_2\lambda tq}
{\eta_1+\eta_2}-\lambda t\right\}(\log x)^{-\frac{3}{4}}
\nonumber \\
&\quad\exp\{2\sqrt{\eta_1\lambda tp}\sqrt{\log x}\}x^{-\eta_1-1}\left(1+O\left((\log x)^{-\frac{1}{2}}\right)\right)
\label{E:ur3}
\end{align}
as $x\rightarrow\infty$, and
\begin{align}
H(t,x)&=\frac{1}{2\sqrt{\pi}}(\eta_2\lambda tq)^{\frac{1}{4}}
\exp\left\{\frac{\eta_1\lambda tp}{\eta_1+\eta_2}-\lambda t\right\}\left(\log\frac{1}{x}\right)^{-\frac{3}{4}}
\nonumber \\
&\quad\exp\left\{2\sqrt{\eta_2\lambda tq}\sqrt{\log\frac{1}{x}}\right\}x^{\eta_2-1}\left(1+O\left(\left(\log\frac{1}{x}\right)^{-\frac{1}{2}}\right)\right)
\label{E:ur4}
\end{align}
as $x\rightarrow 0$.
\end{corollary}

We will next consider the case of the perturbed Heston model where the jump part dominates. 
\begin{theorem}\label{T:glav}
Fix $t> 0$, and suppose $1+\eta_1< A_3$. Then the following asymptotic formula holds for the stock price 
density $D_t$ in the Heston model with double exponential jumps:
\begin{align}
D_t(x)&=\frac{1}{2\sqrt{\pi}}m_{\eta_1}(D_t^{(1)})(\eta_1\lambda tp)^{\frac{1}{4}}\exp\left\{\frac{\eta_2\lambda tq}
{\eta_1+\eta_2}-\lambda t\right\}(\log x)^{-\frac{3}{4}}
\nonumber \\
&\quad\exp\{2\sqrt{\eta_1\lambda tp}\sqrt{\log x}\}x^{-\eta_1-1}\left(1+O\left((\log x)^{-\frac{1}{2}}\right)\right)
\label{E:t1}
\end{align}
as $x\rightarrow\infty$. Here $D_t^{(1)}$ is the density of the random variable $X^{(1)}_t$ defined by 
(\ref{E:razd1}).
\end{theorem}
\begin{theorem}\label{T:nearze}
Fix $t> 0$, and suppose $\widetilde{A}_3>\eta_2-1$. Then the following asymptotic formula holds for the stock 
price density $D_t$ in the Heston model with double exponential jumps:
\begin{align}
D_t(x)&=\frac{1}{2\sqrt{\pi}}m_{-\eta_2}(D_t^{(1)})(\eta_2\lambda tq)^{\frac{1}{4}}
\exp\left\{\frac{\eta_1\lambda tp}{\eta_1+\eta_2}-\lambda t\right\}\left(\log\frac{1}{x}\right)^{-\frac{3}{4}}
\nonumber \\
&\quad\exp\left\{2\sqrt{\eta_2\lambda tq}\sqrt{\log\frac{1}{x}}\right\}x^{\eta_2-1}
\left(1+O\left(\left(\log\frac{1}{x}\right)^{-\frac{1}{2}}\right)\right)
\label{E:zero}
\end{align}
as $x\rightarrow 0$.
\end{theorem}
\begin{remark}\label{R:moment} \rm
The symbols $m_{\eta_1}(D_t^{(1)})$ and $m_{-\eta_2}(D_t^{(1)})$ in formulas (\ref{E:t1}) and (\ref{E:zero}) 
stand for the moments of the marginal density $D_t^{(1)}$ (see (\ref{E:moment})). To get explicit formulas for the 
moments of the Heston density, we may use the equality
\begin{equation}
m_s(D_t^{(1)})=\mathbb{E}\left[\exp\{s\log X_t^{(1)}\}\right]
\label{E:momm}
\end{equation}
and a known explicit formula for the moment generating function of the log-price in the Heston model (see, e.g., formula
(3) in \cite{BRFCU}).
\end{remark}
\begin{remark}\label{R:meani} \rm
Note that formula (\ref{E:t1}) 
becomes meaningless if $1+\eta_1=A_3$. Indeed, if the previous equality holds, then $m_{\eta_1}(D_t^{(1)})=\infty$ 
(use (\ref{E:0})). Similarly, if $\widetilde{A}_3=\eta_2-1$, then $m_{-\eta_2}(D_t^{(1)})=\infty$
(use (\ref{E:est})), and formula (\ref{E:zero}) does not hold.
\end{remark}

\it Proof of Theorem \ref{T:glav}. \rm 
Recall that we denoted by $D^{(1)}_t$ the stock price density in the Heston model. Note that 
the distribution $\mu^{(2)}_t$ of the random variable
$\exp\{T_t\}$ has a singular component at one (see (\ref{E:drvi})). However, we 
can still use a
 formula similar to formula (\ref{E:nach}) to estimate $D_t$. We have
\begin{align}
&D_t(x)=e^{-\lambda t}D^{(1)}_t(x)+D^{(1)}_t\stackrel{M}{\star}H(t,\cdot)(x). 
\label{E:integ1}
\end{align}

Our next goal is to apply Theorem \ref{T:Atg} to characterize the asymptotic behavior of the last term in 
(\ref{E:integ1}). We put $U(x)=D^{(1)}_t(x)$, $\rho=-\eta_1-1$, $l(x)=H_1(t,x)$, $f(x)=H(t,x)$, 
$\sigma=-A_3$, $\tau=\widetilde{A}_3$, and $h(x)=0$. Then, $\sigma<\rho<\tau$. Indeed, the condition 
$1+\eta_1< A_3$ is equivalent to $\sigma<\rho$. In addition, since $\widetilde{A}_3>-1$ (use the integrability
of the function $D_t^{(1)}$ and (\ref{E:est})), we have $\rho<\tau$. Now, taking into account 
Remark \ref{R:const} and Lemma \ref{L:ksv2}, we see that the conditions in the formulation of Theorem 
\ref{T:Atg} hold. It follows that
\begin{align}
D^{(1)}_t\stackrel{M}{\star}H(t,\cdot)(x)&=MD^{(1)}_t(-\eta_1-1)x^{-\eta_1-1}H_1(t,x) 
\left(1+O\left((\log x)^{-\frac{1}{2}}\right)\right)
\label{E:eshnet}
\end{align}
as $x\rightarrow\infty$. Finally, it is not hard to see that (\ref{E:0}), (\ref{E:ur3}), (\ref{E:integ1}), (\ref{E:eshnet}), and the condition $1+\eta_1< A_3$ imply formula (\ref{E:t1}).

This completes the proof of Theorem \ref{T:glav}.

The proof of Theorem \ref{T:nearze} is similar to that of Theorem \ref{T:glav}. It is based on 
Theorem \ref{T:MC0}, (\ref{E:est}), (\ref{E:HH}), Lemma \ref{L:ksv2}, (\ref{E:ur4}), and (\ref{E:integ1}). We leave filling in the details to the interested reader.

We will next explain how the density $D_t$ behaves in the case where the Heston part dominates.
\begin{theorem}\label{T:glavk}
Fix $t> 0$, and suppose $1+\eta_1> A_3$. Then the following asymptotic formula holds for the stock price 
density $D_t$ in the Heston model with double exponential jumps:
\begin{align}
D_t(x)&=\left[e^{-\lambda t}+m_{A_3-1}(H(t,\cdot))\right]
B_1x^{-A_3}\exp\{A_2\sqrt{\log x}\}
(\log x)^{-\frac{3}{4}+\frac{a}{c^2}} \nonumber \\
&\quad\left(1+O\left((\log x)^{-\frac{1}{2}}\right)\right)
\label{E:t1t}
\end{align}
as $x\rightarrow\infty$.  
\end{theorem}
\begin{theorem}\label{T:nearzek}
Fix $t> 0$, and suppose $\widetilde{A}_3<\eta_2-1$. Then the following asymptotic formula holds for the stock price 
density $D_t$ in the Heston model with double exponential jumps:
\begin{align}
D_t(x)&=\left[e^{-\lambda t}+m_{-\widetilde{A}_3-1}(H(t,\cdot))\right]\widetilde{B}_1 x^{\widetilde{A}_3}
\exp\left\{\widetilde{A}_2 \sqrt{\log\frac{1}{x}}\right\} 
\left(\log\frac{1}{x}\right)^{-\frac{3}{4}+\frac{a}{c^2}} \nonumber \\
&\quad\left(1+O\left(\left(\log\frac{1}{x}\right)^{-\frac{1}{2}}\right)\right)
\label{E:zerok}
\end{align}
as $x\rightarrow 0$.
\end{theorem}
\begin{remark}\label{R:moment1} \rm
Recall that the symbols $m_{A_3-1}(\mu_t^{(2)})$ and $m_{-\widetilde{A}_3-1}(\mu_t^{(2)})$ in formulas (\ref{E:t1t}) 
and (\ref{E:zerok}) stand for the moments of the marginal distribution $\mu_t^{(2)}$ (see (\ref{E:moment})). 
To compute the moments appearing in formulas (\ref{E:t1t}) and (\ref{E:zerok}), we can use a formula similar to formula (\ref{E:momm})
and an explicit formula for the moment generating function of the log-price in the exponential jump model with jump amplitudes distributed
according to the asymmetric double exponential law (see, e.g., formula (1) with $b=0$ and $\sigma=0$ in \cite{GMZ}).
\end{remark}
\begin{remark}\label{R:muu} \rm
In the extreme case where $1+\eta_1=A_3$, formula (\ref{E:t1t}) does not hold, since under the previous condition, we
have $m_{-\widetilde{A}_3-1}(\mu_t^{(2)})=\infty$ (use (\ref{E:drvi}), (\ref{E:HH}), and (\ref{E:ur3})). Similarly, 
formula (\ref{E:zerok}) is not valid if
$\widetilde{A}_3=\eta_2-1$, since in this case we have $m_{-\widetilde{A}_3-1}(\mu_t^{(2)})=\infty$
(use (\ref{E:drvi}), (\ref{E:HH}), and (\ref{E:ur4})).
\end{remark}

Theorems \ref{T:glavk} and \ref{T:nearzek} can be derived from formulas
(\ref{E:0S}), (\ref{E:estS}), and (\ref{E:integ1}), using Theorems \ref{T:Atg} and \ref{T:MC0}.

\section{Smile asymptotics in the Heston model with double exponential jumps}\label{S:Hemj}

In order to create a risk-neutral environment, we assume that the following no-arbitrage condition holds for the parameters 
in the perturbed Heston model:
\begin{equation}
\mu=\lambda\left(\frac{q}{\eta_2+1}-\frac{p}{\eta_1-1}\right).
\label{E:riskf}
\end{equation}
Here we take into account that $r=0$. Then the process $\widetilde{X}$ defined by (\ref{D2}) is a martingale (see \cite{G2}, Section 10.8). 
Note that the proof uses the mean-correcting argument (see, e.g., Lemma 10.40 in \cite{G2}, or \cite{S}, pp. 79-80). It will be assumed in the present section that condition (\ref{E:riskf}) holds.

The call and put pricing functions $C$ and $P$ in the Heston model with double exponential jumps are defined by
$C(T,K)=\mathbb{E}\left[(\widetilde{X}_T-K)^{+}\right]$ and
$P(T,K)=\mathbb{E}\left[(K-\widetilde{X}_T)^{+}\right]$,
respectively. In the previous formulas, $T$ is the maturity and $K$ is the strike price. The implied volatility $I(T,K)$, $T> 0$, 
$K> 0$, in the Heston model with double exponential jumps is defined as follows. Given $T$ and $K$, the implied
volatility $I(T,K)$ is equal to the value of the volatility $\sigma=\sigma(T,K)$ in the Black-Scholes model such that
$C(T,K)=C_{BS}(T,K,\sigma)$. Here the symbol $C_{BS}$ stands for the call pricing function in the Black-Scholes model.
In the sequel, the maturity $T$ will be fixed, and we will consider the functions $C$, $P$, and $I$ as functions of only the strike price $K$.

The asymptotic behavior of the implied volatility $I$ in the Heston model with double exponential jumps will be characterized utilizing the asymptotic formulas for the stock price densities provided in Theorems \ref{T:glav}-\ref{T:nearzek}. We will start with the case of large strikes. Analyzing the formulas in Theorems \ref{T:glav} and \ref{T:glavk}, we see that it is important to understand how the implied volatility behaves if the stock price density $D_T$ satisfies the condition
\begin{equation}
D_T(x)=r_1x^{-r_3}\exp\{r_2\sqrt{\log x}\}
(\log x)^{r_4}\left(1+O\left((\log x)^{-\frac{1}{2}}\right)\right)
\label{E:rr1}
\end{equation}
as $x\rightarrow\infty$, where $r_1> 0$, $r_2\ge 0$, $r_3> 2$, and $r_4\in\mathbb{R}$.
\begin{theorem}\label{T:ivo}
Suppose condition (\ref{E:rr1}) holds. Then the following asymptotic formula is valid for the implied volatility:
\begin{align}
&I(K)=\frac{\sqrt{2}}{\sqrt{T}}(\sqrt{r_3-1}-\sqrt{r_3-2})\sqrt{\log \frac{K}{x_0}}+\frac{r_2}{\sqrt{2T}}\left(\frac{1}{\sqrt{r_3-2}}
-\frac{1}{\sqrt{r_3-1}}\right) \nonumber \\
&\quad+\frac{2r_4+1}{2\sqrt{2T}}\left(\frac{1}{\sqrt{r_3-2}}-\frac{1}{\sqrt{r_3-1}}\right)
\frac{\log\log\frac{K}{x_0}}{\sqrt{\log\frac{K}{x_0}}} 
\nonumber \\
&\quad+\left[\frac{1}{\sqrt{2T}}\log\frac{\sqrt{r_3-1}\sqrt{r_3-2}(\sqrt{r_3-1}-\sqrt{r_3-2})}{2\sqrt{\pi}r_1}
+\frac{r_2^2}{4\sqrt{2T}}\left(\frac{1}{(r_3-2)^{\frac{3}{2}}}-\frac{1}{(r_3-1)^{\frac{3}{2}}}\right)\right]\frac{1}{\sqrt{\log\frac{K}{x_0}}}
\nonumber \\
&\quad+\frac{r_2(2r_4+1)}{4\sqrt{2T}}\left(\frac{1}{(r_3-2)^{\frac{3}{2}}}-\frac{1}{(r_3-1)^{\frac{3}{2}}}\right)\frac{\log\log\frac{K}{x_0}}{\log\frac{K}{x_0}}
+O\left(\frac{1}{\log\frac{K}{x_0}}\right)
\label{E:rr2}
\end{align}
as $K\rightarrow\infty$.
\end{theorem}

\it Proof. \rm For the sake of simplicity, we assume $x_0=1$. The proof in the general case is similar.

It follows from (\ref{E:rr1}), Corollary 7.13 in \cite{G2}, and Theorem 8.10 in \cite{G2} that as $K\rightarrow\infty$,
\begin{equation}
C(K)=\frac{r_1}{(r_3-1)(r_3-2)}(\log K)^{r_4}\exp\{r_2\sqrt{\log K}\}K^{2-r_3}
\left(1+O\left(\frac{1}{\sqrt{\log K}}\right)\right).
\label{E:si}
\end{equation}
Therefore, as $K\rightarrow\infty$,
\begin{align}
\log\frac{1}{C(K)}&=\log\frac{(r_3-1)(r_3-2)}{r_1}-r_4\log\log K-r_2\sqrt{\log K}
+(r_3-2)\log K \nonumber \\
&\quad+O\left(\frac{1}{\sqrt{\log K}}\right).
\label{E:logg}
\end{align}
Moreover, the mean value theorem and (\ref{E:logg}) imply that
$$
\log\log\frac{1}{C(K)}=\log\log K+\log(A_3-2)+O\left(\frac{1}{\sqrt{\log K}}\right)
$$
as $K\rightarrow\infty$. Next, using Theorem 9.16 in \cite{G2} with $\lambda=r_3-2$ and $\Lambda(K)=\sqrt{\log K}$,
we get
\begin{equation}
I(K)=\frac{\sqrt{2}}{\sqrt{T}}\left[\sqrt{(r_3-1)\log K+L(K)}-\sqrt{(r_3-2)\log K+L(K)}\right]
+O\left(\frac{1}{\log K}\right)
\label{E:logg1}
\end{equation}
as $K\rightarrow\infty$, where
\begin{align*}
L(K)&=-r_2\sqrt{\log K}-\left(r_4+\frac{1}{2}\right)\log\log K+\log\frac{(r_3-1)(r_3-2)}{r_1} \\
&\quad-\log(r_3-2)+\log\frac{\sqrt{r_3-1}-\sqrt{r_3-2}}{2\sqrt{\pi}{r_3-1}}.
\end{align*}
It follows from (\ref{E:logg1}) that
\begin{align}
I(K)&=\frac{\sqrt{2}}{\sqrt{T}}\left[\sqrt{r_3-1}\sqrt{\log K}\sqrt{1+\frac{L(K)}{(r_3-1)\log K}}
-\sqrt{r_3-2}{\sqrt{\log K}}\sqrt{1+\frac{L(K)}{(r_3-2)\log K}}\right] \nonumber \\
&\quad+O\left(\frac{1}{\log K}\right)
\label{E:logg2}
\end{align}
as $K\rightarrow\infty$. Now, using the formula
$\sqrt{1+h}=1+\frac{1}{2}h-\frac{1}{8}h^2+O\left(h^3\right)$, $h\rightarrow 0$,
in (\ref{E:logg2}), and making simplifications, we obtain formula (\ref{E:rr2}) with $x_0=1$.

This completes the proof of Theorem \ref{T:ivo}.

The next theorem characterizes the asymptotic behavior of the implied volatility at large strikes in the Heston model 
with double exponential jumps.
\begin{theorem}\label{T:hdej}
Let $T> 0$, and suppose $1+\eta_1< A_3$. Then formula (\ref{E:rr2}) holds with
$$
r_1=\frac{1}{2\sqrt{\pi}}m_{\eta_1}(D_T^{(1)})(\eta_1\lambda Tp)^{\frac{1}{4}}\exp\left\{\frac{\eta_2\lambda Tq}
{\eta_1+\eta_2}-\lambda T\right\},
$$ 
$$
r_2=2\sqrt{\eta_1\lambda Tp},\quad r_3=\eta_1+1,\quad\mbox{and}\quad r_4=-\frac{3}{4}.
$$
On the other hand, if $1+\eta_1> A_3$, then formula (\ref{E:rr2}) holds with
$$
r_1=\left[e^{-\lambda T}+m_{A_3-1}(H(T,\cdot))\right]B_1,
$$ 
$$
r_2=A_2,\quad r_3=A_3,\quad\mbox{and}\quad r_4=-\frac{3}{4}+\frac{a}{c^2}.
$$
\end{theorem}

\it Proof. \rm Theorem \ref{T:hdej} follows from Theorems \ref{T:glav}, \ref{T:glavk}, and \ref{T:ivo}.
\begin{remark}\label{R:reem}
In the paper \cite{GL} of K. Gao and R. Lee, an asymptotic formula with four terms and an error estimate of order 
\begin{equation}
O\left((\log K)^{-\frac{3}{4}}\right),\quad K\rightarrow\infty,
\label{E:neo}
\end{equation} 
was found for the implied volatility in the negatively correlated Heston model (see \cite{GL}, Corollary 8.1). Using (\ref{E:0}) and Theorem \ref{T:ivo}, we can obtain a sharper asymptotic formula with five terms and an error estimate of order $O\left((\log K)^{-1}\right)$ as $K\rightarrow\infty$. The fifth term in this formula is of the form 
$c\frac{\log\log K}{\log K}$. The previous expression tends to zero faster than the expression in (\ref{E:neo}). The reason why the formula obtained in \cite{GL} contains a weaker error estimate is the following. In the proof of their result, Gao and Lee used formula (4.2) in \cite{FGGS}, which is an asymptotic formula for the call pricing function in the Heston model with a relative error estimate $O\left((\log K)^{-\frac{1}{4}}\right)$. However, formula (4.2) in \cite{FGGS} contains a typo. More precisely, the power $-\frac{1}{4}$ in the error estimate in that formula can be replaced by the power $-\frac{1}{2}$. Indeed, it suffices to integrate the expressions in formula (4.1) in \cite{FGGS} twice. 
Formula (4.1) is an asymptotic formula for the stock price density in the negatively correlated Heston model, 
containing a correct relative error estimate $O\left((\log x)^{-\frac{1}{2}}\right)$. Note that in the presentation of the results from \cite{FGGS} in the book \cite{G2}, the asymptotic formula for the call pricing function in the negatively correlated Heston model contains a correct error term (see formula (8.28) in \cite{G2}).
\end{remark}
\begin{remark}\label{R:gerhold}
The asymptotic behavior of the implied volatility at large strikes in Kou's model was studied in \cite{Z} and \cite{GMZ}. Since Kou's model is the Black-Scholes model with double exponential jumps, the jump part always dominates. Indeed, the decay of the stock price density in the Black-Scholes model is log-normal, while the density of the exponential L\'{e}vy part
of Kou's model decays as a regularly varying function. The authors of \cite{GMZ} obtain an asymptotic formula for the implied volatility with four terms an an error estimate of order $O\left((\log K)^{-\frac{3}{4}}\right)$ as in \cite{GL}. It is not hard to obtain a similar expansion with five terms and an error estimate of order $O\left((\log K)^{-1}\right)$ as $K\rightarrow\infty$, using Theorems \ref{T:Atg}, \ref{T:mainre}, and \ref{T:ivo} established in the present paper. 
\end{remark}

Our next goal is to characterize the asymptotic behavior of the implied volatility at small strikes. Here we
borrow various ideas used in Section 9.7 of \cite{G2}. For the sake of shortness, we again assume that $x_0=1$.
Set $G(K)=KP\left(K^{-1}\right)$. Then $G$ is a call pricing function. The corresponding marginal densities are as follows: 
\begin{equation}
\widetilde{D}_T(x)=x^{-3}D_T\left(x^{-1}\right)
\label{E:magr1}
\end{equation}
(see Remark 9.20 in \cite{G2}). Now, using Theorems \ref{T:nearze} and \ref{T:nearzek}, formula (\ref{E:magr1}),
and the fact that $I(K)=I_G\left(K^{-1}\right)$ (see Lemma 9.23 in \cite{G2}), we obtain the following assertion.
\begin{theorem}\label{T:kon}
Let $T> 0$, and suppose $\widetilde{A}_3>\eta_2-1$. Then the following formula holds for the implied volatility in the Heston model with double exponential jumps:
\begin{align}
&I(K)=\frac{\sqrt{2}}{\sqrt{T}}(\sqrt{s_3+1}-\sqrt{s_3})\sqrt{\log\frac{x_0}{K}}
+\frac{s_2}{\sqrt{2T}}\left(\frac{1}{\sqrt{s_3}}-\frac{1}{\sqrt{s_3+1}}\right)
\nonumber \\
&\quad+\frac{2s_4+1}{2\sqrt{2T}}\left(\frac{1}{\sqrt{s_3}}-\frac{1}{\sqrt{s_3+1}}\right)
\frac{\log\log\frac{x_0}{K}}{\sqrt{\log\frac{x_0}{K}}} \nonumber \\
&\quad+\left[\frac{1}{\sqrt{2T}}\log\frac{\sqrt{s_3+1}\sqrt{s_3}(\sqrt{s_3+1}-\sqrt{s_3})}
{2\sqrt{\pi}s_1}+\frac{s_2^2}{4\sqrt{2T}}\left(\frac{1}{s_3^{\frac{3}{2}}}-\frac{1}{(s_3+1)^{\frac{3}{2}}}
\right)\right]\frac{1}{\sqrt{\log\frac{x_0}{K}}} \nonumber \\
&\quad+\frac{s_2(2s_4+1)}{4\sqrt{2T}}\left(\frac{1}{s_3^{\frac{3}{2}}}-\frac{1}{(s_3+1)^{\frac{3}{2}}}
\right)\frac{\log\log\frac{x_0}{K}}{\log\frac{x_0}{K}}+O\left(\frac{1}{\log\frac{x_0}{K}}\right)
\label{E:magr2}
\end{align}
as $K\rightarrow 0$. In (\ref{E:magr2}),
$$
s_1=\frac{1}{2\sqrt{\pi}}m_{-\eta_2}\left(D_T^{(1)}\right)(\eta_2\lambda Tq)^{\frac{1}{4}}
\exp\left\{\frac{\eta_1\lambda Tp}{\eta_1+\eta_2}-\lambda T\right\},
$$
$$
s_2=2\sqrt{\eta_2\lambda Tq},\quad s_3=\eta_2,\quad\mbox{and}\quad s_4=-\frac{3}{4}.
$$

On the other hand, if $\widetilde{A}_3<\eta_2-1$, then (\ref{E:magr2}) holds with
$$
s_1=\left[e^{-\lambda T}+m_{A_3-1}(H(T,\cdot))\right]\widetilde{B}_1,
$$
$$
s_2=\widetilde{A}_2,\quad s_3=\widetilde{A}_3+1,\quad\mbox{and}\quad s_4=-\frac{3}{4}+\frac{a}{c^2}.
$$
\end{theorem}

\section{More applications}\label{S:moor}
The example discussed in this section is a stochastic stock price model that is the mixture of the Heston model with a special exponential L\'{e}vy model. The log-price process in the perturbing model is the L\'{e}vy process with normal inverse Gaussian marginals (the NIG process). We will characterize the asymptotic behavior of the implied volatility at large strikes in the perturbed Heston model mentioned above. The behavior at the small strikes can be characterized similarly.

The normal inverse Gaussian distribution and the NIG process were introduced by O. Barndorff-Nielsen in \cite{BNa} 
and \cite{BN}, respectively (see also \cite{BNb}). For the sake of simplicity, we will consider only symmetric centered NIG-processes. 
The general case can be dealt with similarly. Let $\alpha> 0$ and $\delta> 0$, let $W^{(\alpha)}_t=W_t+\alpha t$, $t\ge 0$, 
be Brownian motion with drift, and let $A$ be the inverse Gaussian process given by
$A_t=\inf\left\{s> 0:W^{(\alpha)}_s=\delta t\right\}$.
Consider also an independent standard Brownian motion: $\widetilde{W}_t$, $t\ge 0$. Then the NIG-process is
defined by $Y_t=\widetilde{W}_{A_t}$, $t\ge 0$.
The parameter $\alpha$ controls the tail heaviness of marginal distributions, while $\delta$ is the scale 
parameter.

Let us consider a mixed model  
$X_t=X^{(1)}_tX^{(2)}_t$, $t\ge 0$,
where $X^{(1)}$ is the price process in the Heston model defined in (\ref{E:Heston1}), while  $X^{(2)}=\exp\{Y_t\}$. 
As before, we denote by $D^{(k)}_t$ the distribution density of the random variable $X^{(k)}_t$, $k=1,2$, 
and by ${\widetilde D}^{(2)}_t$ the density of $Y_t$. 

There exists a closed-form expression for the density $D^{(2)}_t$. The modified Bessel function $K_1$ of the 
third kind will be needed in the sequel. This function is defined by
$$
K_1(z)=\frac{1}{2}\int_0^{\infty} \exp\left\{-\frac{z}{2}(u+\frac{1}{u})\right\}du.
$$
Denote by $(\gamma, 0,\nu)$ the L\'{e}vy triplet associated with the process $Y$. It is known that 
$$
\gamma=\frac{2\alpha\delta}{\pi}\int_0^1 K_1(\alpha x)dx
$$
and 
$$
\nu(dy)=\frac{\alpha\delta}{\pi}\frac{K_1 (\alpha |y|)}{|y|}dy.
$$
The following formula holds for the density $\widetilde{D}^{(2)}_t$:
$$
{\widetilde D}^{(2)}_t(y)=k(t)\frac{K_1 (\alpha\sqrt{y^2+\delta^2 t^2})}{\sqrt{y^2+\delta^2 t^2}},
$$
where 
$$
k(t)=\frac{\alpha\delta t e^{\alpha\delta t}}{\pi}.
$$
Therefore, 
$$
D^{(2)}_t (x)=\frac{k(t)}{x} \frac{K_1 (\alpha\sqrt{(\log x)^2+\delta^2 t^2})}{\sqrt{(\log x)^2+\delta^2 t^2}}, 
$$
for all $x> 0$. The previous formulas can be found in \cite{S}. 
It is known that 

$$
K_1(z)=\sqrt{\frac{\pi}{2z}} e^{-z}\left(1+O\left(\frac{1}{z}\right)\right)
$$
as $z\rightarrow\infty$ (see formula 9.7.2 in \cite{AS}).
It follows that
\begin{equation}
D^{(2)}_t (x)=\frac{k(t)}{x}\sqrt{\frac{\pi}{2\alpha}} 
\frac{\exp\left\{-\alpha \sqrt{(\log x)^2+\delta^2 t^2}\right\}}{((\log x)^2+\delta^2t^2)^{\frac{3}{4}}}
\left(1+O\left(\frac{1}{\log x}\right)\right)
\label{E:s1}
\end{equation}
as $x\rightarrow\infty$.

Since for every $A> 0$ and $\alpha> 0$,
$$
\frac{1}{\left((\log x)^2+A\right)^{\frac{3}{4}}}=\frac{1}{\left((\log x)^2\right)^{\frac{3}{4}}}
\left(1+O\left(\frac{1}{\log x}\right)\right)
$$
and
$$
\exp\left\{-\alpha \sqrt{(\log x)^2+A}\right\}=\exp\left\{-\alpha \sqrt{(\log x)^2}\right\}
\left(1+O\left(\frac{1}{\log x}\right)\right)
$$
as $x\rightarrow\infty$, formula (\ref{E:s1}) implies that
\begin{equation}
D^{(2)}_t (x)=k(t)\sqrt{\frac{\pi}{2\alpha}} 
x^{-\alpha-1}(\log x)^{-\frac{3}{2}}\left(1+O\left(\frac{1}{\log x}\right)\right)
\label{E:s2}
\end{equation}
as $x\rightarrow\infty$.

It is not hard to see, using formulas (\ref{E:s2}) and (\ref{E:0S}), that
for $A_3<\alpha+1$, the Heston part of the mixed model dominates, while for $\alpha+1< A_3$, the NIG part dominates. 
Our next goal is to apply Theorem \ref{T:ivo} to the Heston$+$NIG model. The no-arbitrage condition for this model is the following:
\begin{equation}
\alpha\ge 1\quad\mbox{and}\quad \mu=\delta(\sqrt{\alpha^2-1}-\alpha),
\label{E:noar}
\end{equation}
where $\mu$ is the drift parameter in the Heston model. Recall that we assume that $r=0$. Condition (\ref{E:noar})
can be obtained, using the mean-correcting argument (see the references in the beginning of Section \ref{S:Hemj}) and an explicit formula
for the characteristic function of the NIG distribution (see \cite{BN}, see also \cite{S}, Section 5.39).
\begin{theorem}\label{T:fini}
Suppose the no-arbitrage condition in (\ref{E:noar}) holds, and let $A_3<\alpha+1$. Then formula (\ref{E:rr2}) is valid for the 
implied volatility in the Heston$+$NIG model
with 
$$
r_1=m_{A_3-1}\left(D^{(2)}_T\right)B_1,
$$ 
$r_2=A_2$, $r_3=A_3$, and $r_4=-\frac{3}{4}+\frac{a}{c^2}$.
In addition, if $\alpha+1< A_3$, then formula (\ref{E:rr2}) is valid with
$$
r_1=k(t)\sqrt{\frac{\pi}{2\alpha}}m_{\alpha}\left(D^{(1)}_T\right), 
$$
$r_2=0$, $r_3=\alpha+1$, and $r_4=-\frac{3}{2}$.
\end{theorem}

Theorem \ref{T:fini} follows from (\ref{E:0S}), (\ref{E:s2}), Theorem \ref{T:Atg}, and Theorem \ref{T:ivo}. A similar theorem can be obtained
in the case where $K\rightarrow 0$. We leave the formulation and the proof of such a theorem as an exercise for the interested reader.
\begin{remark}\label{R:univerk}
The methods developed in the present paper are rather universal. They can be used to approximate the stock price density and the implied volatility in many mixed stochastic models. For instance, we can replace the Heston model with jumps by the Stein-Stein model with jumps (see \cite{G2} for the discussion of the asymptotic behavior of the stock price density in the Stein-Stein model), and also use jump processes different from the double exponential process or the NIG process. We only need to know appropriate asymptotic formulas with error estimates for the marginal distributions of the jump process, and such formulas are often available.
\end{remark}
\section{Acknowledgments}
The authors thank R. Lee and S. De Marco for valuable comments, and also S. Gerhold for providing the references \cite{E} 
and \cite{W}.

\end{document}